\documentclass[twocolumn,twocolappendix]{aastex631}
\usepackage[utf8x]{inputenc}
\usepackage{amsmath,amsfonts,amssymb,mathrsfs,bm,enumerate,graphicx}
\usepackage[T1]{fontenc}
\usepackage[all]{hypcap}
\usepackage{ulem}

\usepackage{ae,aecompl}
\hypersetup{breaklinks,colorlinks,citecolor=blue,linkcolor=magenta}

\newcommand{\ppderiv}[2]{\frac{\partial #1}{\partial #2}}

\newcommand{\del}{\mathbf{\nabla}}

\newcommand{\avg}[1]{\left\langle #1 \right\rangle}

\newcommand{\uvec}[1]{\hat{{\bm #1}}}

\newcommand{\be}{\begin{eqnarray}}
\newcommand{\ee}{\end{eqnarray}}

\newcommand{\bmc}[1]{\textcolor[rgb]{.5, 0, 0}{1}}
\newcommand{\rev}[1]{#1}

\newcommand{\bmat}{\begin{pmatrix}}
\newcommand{\emat}{\end{pmatrix}}

\renewcommand{\vec}[1]{{\bf #1}}

\newcommand{\isolatedp}[1]{\citep[#1][]{Tang.2017,Moody.2019,Munoz.2019,Munoz.2020,Duffell.2020,Tiede.2020,Zrake.2020knc,DOrazio.2021,Dittmann.2021,Dittmann.2022}}

\shorttitle{Contracting and Expanding BBHs in 3D AGN Disks}
\shortauthors{Dempsey, Li, Mishra, \& Li}

\begin{document}

\title{Contracting and Expanding Binary Black Holes in 3D Low-Mass AGN Disks: The Importance of Separation}
\author[0000-0001-8291-2625]{Adam M. Dempsey}
 \affiliation{X-Computational Physics Division, Los Alamos National Laboratory, Los Alamos, NM 87545, USA}
 \affiliation{Theoretical Division, Los Alamos National Laboratory, Los Alamos, NM 87545, USA}
 \author[0000-0003-3556-6568]{Hui Li}
 \affiliation{Theoretical Division, Los Alamos National Laboratory, Los Alamos, NM 87545, USA}
 \author[0000-0003-0271-3429 ]{Bhupendra Mishra}
 \affiliation{Theoretical Division, Los Alamos National Laboratory, Los Alamos, NM 87545, USA}
\author[0000-0002-4142-3080]{Shengtai Li}
\affiliation{Theoretical Division, Los Alamos National Laboratory, Los Alamos, NM 87545, USA}

\correspondingauthor{Adam M. Dempsey}
\email{adempsey@lanl.gov}

\begin{abstract}

LIGO/Virgo has detected several binary black hole (BBH) merger events that may have originated in the accretion disks of Active Galactic Nuclei (AGN). 
These events require individual black hole masses that fall within the pair instability supernova mass gap, and therefore these black holes may have been grown from hierarchical mergers. 
AGN disks are a prime environment for hierarchical mergers, and thus a potential location for the progenitors of BBH gravitational wave events. 
Understanding how a BBH embedded in an AGN disk interacts with the surrounding environment is thus crucial for determining if this interaction can lead to its merger.
However, there are few high fidelity simulations of this process, and almost all are two-dimensional. 
We present the results from 3D, high-resolution, local shearing-box simulations of an embedded BBH interacting with an AGN disk. 
In these first simulations of their kind, we focus on determining the mass accretion rate and the orbital evolution rate at different BBH separations. 
We find that circular, equal-mass BBHs with separations greater than 10\% of their Hill radius contract while accreting at a super-Eddington rate.
At smaller separations, however, our 3D simulations find that BBHs expand their orbits.
This result suggests that it may be difficult for an AGN disk to push a BBH to merger, but we discuss several mechanisms, including MHD turbulence and radiative and mechanical feedback, that could alleviate this difficulty.

\end{abstract}

\section{Introduction} \label{sec:intro}

Active Galactic Nucleus (AGN) disks around super massive black holes (SMBH) are one of the pathways to producing merging black holes (BHs). 
BHs in these disks may grow via super-Eddington accretion and repeated mergers to eventually produce heavyweight binary BH (BBH) gravitational wave (GW) events \citep[see e.g.][for a review of hierarchical mergers]{Gerosa.2021}.
At least two recent GW events may have come from merging BBHs embedded in AGN disks:
\begin{itemize}
    \item GW190521 was a merger of two BHs with total mass $\sim 150 M_\odot$ and with at least one BH more massive than $\sim 65 M_\odot$ \citep{Abbott.2020a,Abbott.2020b}.
This GW event was accompanied by an electromagnetic counterpart from a known AGN observed by the ZTF \citep{Graham.2020} that was spatially coincident with the GW localization \citep{Chen.2020}. 
    \item GW170817A was an additional GW trigger on the day of the detected neutron star merger GW1701817 that was found by an independent analysis after the fact \citep{Zackay.2019}. 
The large mass and aligned spin has been argued as having come from a merger in an AGN disk \citep{Gayathri.2019}. 
\end{itemize}

While the association of these GW events with BBHs in AGN disks is still speculative at best \citep{Ashton.2020,Palmese.2021}, it opens up the possibility for a new BBH formation channel that has not been thoroughly explored. 
Much work on merging BBHs in AGN disks has been carried out with simplified models that are typically in the guise of population synthesis calculations \citep{McKernan.2012,McKernan.2014,McKernan.2018,McKernan.2019,McKernan.2020,McKernan.2021,Yang.2019a,Yang.2019b,Secunda.2019,Secunda.2020,Tagawa.2020,Tagawa.2021}. 
These models, while important to develop a statistical sense of the overall problem and its connection to observations, vastly simplify the interaction of the BBH with the AGN disk. 
By and large, the binary-disk interaction prescriptions used in these works were developed and tuned to the conditions found in circumstellar/protoplanetary disks. 
These environments can be wildly different from AGN environments, and in particular the picture of a stellar binary interacting with a large viscous accretion disk is not necessarily a direct analogue of a BBH embedded in an accretion disk centered on a SMBH.
To date, there have only been a few 2D \citep{Baruteau.2010,Li.2021,Li.2022,Lai.2022} and 3D \citep{Kaaz.2021crx} hydrodynamical simulations of BBHs interacting with an AGN disk. 

Conventional wisdom from nearly 15 years ago was that binary-disk interaction should result in contraction \citep{MacFadyen.2006}. 
But, recent high resolution simulations have found the opposite -- binaries tend to expand \isolatedp{e.g.,}. 
The key missing ingredient in the early studies was the interaction of the binary with the circumsingle disks (CSDs) around each binary component, which has since been found to increase the binary's angular momentum.
Taking these results at face value, one may be worried that BBHs in AGN disks should never merge. 
But, binary expansion may be a special outcome of circular, equal-mass binaries. 
Indeed, simulations of eccentric and non-equal mass binaries find that sufficiently eccentric binaries contract \citep{Munoz.2019,DOrazio.2021}, as do sufficiently low-mass ratio binaries \citep{Duffell.2020,Dempsey.2021}.

Placing a binary inside of another disk introduces even more avenues for removing the binary energy and angular momentum.
Our previous works have explored a few possibilities. 
In \citet{Li.2021}, we found with high-resolution 2D simulations that embedded, retrograde, equal-mass, circular BBHs contract, whereas prograde BBHs expand if the gravitational softening of each BH is small enough to sufficiently resolve the CSDs.
Further work in \citet{Li.2022} has found that prograde binaries can contract if each BH is able to heat its entire CSD by at least a factor of $\sim 3$ over the AGN disk temperature.
But, these initial works have only explored a small subset of the parameter space and, in particular, were limited to 2D simulations.

In this work, we expand upon our previous simulations and study how an embedded BBH interacts with the AGN disk in 3D. 
The layout of this paper is as follows.
In Section \ref{sec:background}, we present a background on AGN disk structure and how it relates to the BBH properties. 
We present our numerical method in Section \ref{sec:method}, and the results of our 3D simulations in Section \ref{sec:results}.
We discuss our results in Section \ref{sec:discussion}, and present our conclusions in Section \ref{sec:conclusion}.

\section{A Global Picture of BBH-disk interaction} \label{sec:background}

In this Section we provide a basic overview of the problem of a BBH interacting with an AGN disk. 
We tackle the problem from the point of view of a hierarchy of length scales and identify a few key parameters that control how these binaries accrete and how they evolve.

\subsection{The AGN Scale} \label{sec:agn_scale}

AGN disks can span more than $\gg 10^7 R_g$ in size, where $R_g = G M_{\rm smbh}/c^2$ is the gravitational radius of a SMBH with mass $M_{\rm smbh}$. 
Unlike their stellar counterparts, models of AGN disks typically require a treatment of radiation as it can be important dynamically.
In particular, the opacity profile of the disk plays a critical role in determining its thickness, stability, and possible migration traps for migrating objects \citep{Sirko.2003,Thompson.2005,Bellovary.2016,Dittmann.2019}.

Models of AGN disks determine the radial profiles of the gas density, pressure, vertical scale height, temperature, and opacity. 
In Figure \ref{fig:agn}, we show a collection of AGN models taken from \citet{Sirko.2003}, \citet{Thompson.2005}, and \citet{Dittmann.2019}\footnote{These particular \citet{Sirko.2003} and \citet{Thompson.2005} profiles are taken from Figure 1 of \citet{Bellovary.2016}. Additionally, the \citet{Dittmann.2019} models apply the \citet{Thompson.2005} model to high-redshift, low-mass AGN.}.
For each model, we show the radial profiles of aspect ratio ($H/R$), surface density ($\Sigma$), and temperature ($T$). 
These models all assume that radiation pressure is important, and that heating due to radial accretion in the disk balances vertical cooling. 
Additionally, all models assume that there is some form of feedback that helps keep the disk marginally gravitationally stable.  
Both the \citet{Sirko.2003} and \citet{Thompson.2005} models are for a $10^8 M_\odot$ SMBH, while the \citet{Dittmann.2019} models are for a lower mass $4 \times 10^6 M_\odot$ SMBH. 

Despite the different AGN masses, all models show the same general structure. 
AGN disks are characterized by a ``bowl" shaped $H/R$ profile: there is a thin middle region around $10^3-10^4 R_g$ surrounded by thicker interior and exterior regions. 
The inner regions are typically gravitationally stable and follow a \citet{Shakura.1973} $\alpha$-disk like structure, while the outer disks may be radiation pressure dominated and maintain $Q\sim 1$ via feedback processes. 
The \citet{Thompson.2005} model, in particular, has a very thin middle region $H/R \sim 10^{-3}$ compared to the $H/R\sim 0.01$ values found in the \citet{Sirko.2003} and \citet{Dittmann.2019} models. 
The AGN disks of \citet{Thompson.2005} are also generally cooler and less dense.

Attempts to model AGN disks as $\alpha$-disks have found that these disks can be unstable to viscous, gravitational, and/or thermal instabilities \citep[e.g.,][]{Lightman.1974,Piran.1978}.
Ways to circumvent this include strong vertical magnetic fields \citep[e.g.,][]{Pariev.2003} or enhanced radiation pressure from outflows sourced by a population of stars or BHs \citep[e.g.,][]{Sirko.2003,Thompson.2005,Dittmann.2019}. 
In fact, BHs formed in the very outer regions of the AGN disk may even supply the progenitors to BBHs at smaller radii \citep[e.g.,][]{Dittmann.2019,Cantiello.2021}.

\subsection{The Hill Radius} \label{sec:hill_scale}

\begin{figure}
\centering
\includegraphics[width=.48\textwidth]{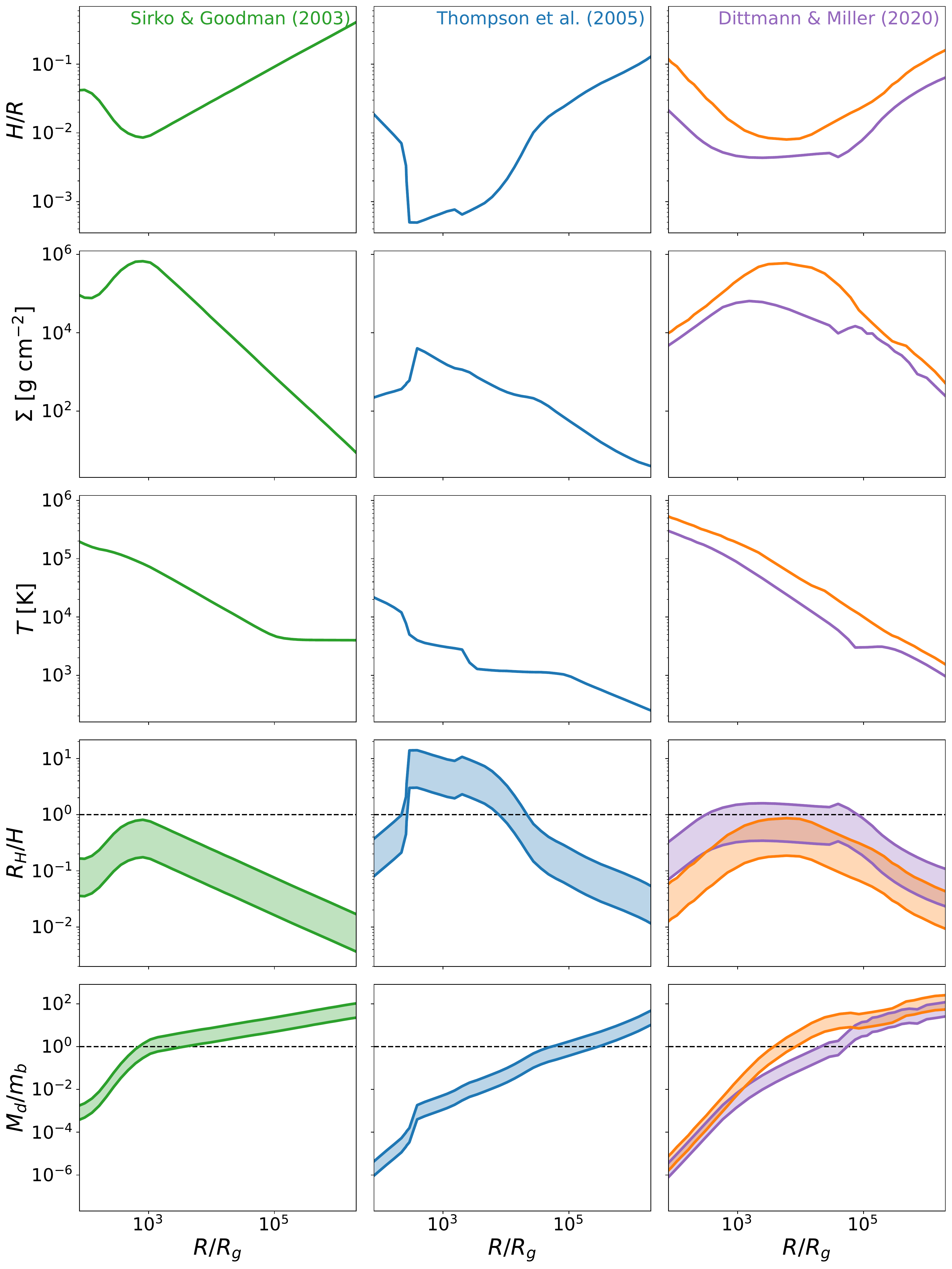}
\caption{
Comparison between four AGN disk models taken from \citet[][left]{Sirko.2003}, \citet[][middle]{Thompson.2005}, and \citet[][right]{Dittmann.2019}. 
Note that we show two models from \citet{Dittmann.2019}. 
From \rev{the first to third rows} we plot the radial profiles of aspect ratio ($H/R$), surface density ($\Sigma$), and temperature ($T$). 
In the \rev{fourth and fifth rows} we show the plausible ranges of $R_H/H$ and the disk-to-binary mass for $m_b/M_{\rm smbh} = [10^{-8},10^{-6}]$. 
At $R \sim 10^3-10^4 R_g$, $R_H/H \lesssim 1$ and $M_d/m_b \lesssim 1$, where $M_d=4\pi R_H^2 \Sigma$ is a measure of the local disk mass.
The \citet{Thompson.2005} model generally has larger $R_H/H$ and smaller $M_d/m_b$ in this range as the AGN disk is a factor of $\sim 10$ times thinner than the other models. 
}
\label{fig:agn}
\end{figure}

We consider a BBH (of mass $m_b$) embedded in the AGN disk at some orbital radius, $R_0$. 
The critical length scale associated with the binary is its Hill radius, $R_H = (m_b/(3 M_{\rm smbh}))^{1/3} R_0$.
If we consider, for example, a $60 M_\odot$ binary in orbit around a $10^8 M_\odot$ SMBH at $R_0 = 10^4 R_g$, the Hill radius will be $R_H \sim 0.006 R_0$, which could be smaller than the vertical thickness of the disk if $H \gtrsim 0.01 R_0$ (as Figure \ref{fig:agn} suggests). 
In terms of the gravitational radius of each stellar mass BH, the Hill radius is $\sim 10^8 r_g$, where $r_g =  G m_{\rm bh}/c^2$ and $m_{\rm bh}$ is the individual BH mass. 
This is well outside the radius where GR effects are important. 

The ratio of the Hill radius to the disk scale height is an important quantity. 
If $R_H \gg H$, then the BBH may open a gap in the disk \citep{Crida.2006}, and the flow into the BBH Hill sphere may be treated in 2D. 
On the other hand if $R_H \ll H$, then the flow onto the binary is Bondi-like and 3D \citep{Kaaz.2021crx}. 
There is also the marginal case where $R_H \lesssim H$.
Here, we would expect there to be sizable CSDs around each BH. 
The structure of the CSDs plays a critical role in setting how accretion onto each BH proceeds. 
And, as previously mentioned, determining how the binary orbit evolves depends critically on the CSDs. 

The shaded regions in the fourth row of Figure \ref{fig:agn} show a range of plausible values for $R_H/H$ for BBHs with mass ratios $m_b/M_{\rm smbh}=[10^{-8},10^{-6}]$.  
In the $R\sim 10^3 - 10^4 R_g$ regions, $R_H/H \sim 0.1 - 1$ in both the \citet{Sirko.2003} and \citet{Dittmann.2019} models, whereas the \citet{Thompson.2005} models have $R_H/H \gg 1$ due to the very small $H$ there. 
In the thicker regions of the disk, $R_H/H \ll 1$, which implies that accretion onto BBHs here may proceed in a more Bondi-like fashion \citep{Kaaz.2021crx}.

In addition to $R_H/H$, another key parameter controlling the evolution of an embedded BBH is how much gas is in its immediate vicinity. 
More material in the BBH Hill sphere results in both faster orbital evolution rates and faster mass growth through accretion. 
To estimate this parameter, we show with the shaded regions in the bottom row of Figure \ref{fig:agn} the ratio of the local disk mass, $4\pi R_H^2 \Sigma$, to the BBH mass for the same $m_b/M_{\rm smbh}$ values as before.
\rev{All models agree that the local disk becomes more massive further out in the AGN disk, but differ on what radius this mass equals the BBH mass.
The BBH is typically more massive than its surroundings at a radius of $\sim 10^5 R_g$ in \citet{Thompson.2005}; $\sim 3\times 10^3 - 3 \times 10^4 R_g$ for the \citet{Dittmann.2019} models; and $\sim 10^3 R_g$ in the \citet{Sirko.2003} disk.}
\rev{Interior to these radii,} the time scale for the mass and orbital properties of the BBH to change may be ``slow" in the sense that we would expect the gas around the BBH to be in quasi-steady-state \citep[see the discussions of steady-state in e.g.,][]{Dempsey.20204f,Dempsey.2021}.

\subsection{The Binary Scale} \label{sec:binary_scale}
The next scale down in the hierarchy is the size of the BBH orbit. 
This can range anywhere from a few $r_g$ to a fraction of $R_H$. 
However, binaries with separations approaching a large fraction of $R_H$ ($a_b \gtrsim 0.4 R_H$) are known to evolve chaotically and are likely disrupted by the tidal force of the SMBH \citep[][and see Figure \ref{fig:nbody} below]{Eggleton.1995,Mardling.2001}.
In the limit $a_b \ll R_H$, the BHs are surrounded by a large circumbinary disk (CBD) from which they feed.
Upon their formation, though, the BHs may be separated on much wider scales. 
In this limit, each BH is surrounded by a large CSD that extends out to $\sim 0.4 a_b$ \citep{Artymowicz.1994}, with the remaining space between $a_b$ and $R_H$ taken up by the CBD.

When the binary separation is on the order of $R_H$, the fact that the binary is embedded in a larger scale accretion disk is an important distinction compared to an isolated binary. 
In particular, the size and role of the CBD is very different in this situation compared to the large, viscously controlled CBD of an isolated binary \citep[as pointed out in][]{Li.2021}. 

\rev{Even when $a_b \ll R_H$, embedded binaries still have important differences with respect to their isolated counterparts. In particular, recent studies have highlighted the importance of retrograde orbits \citep{Li.2021}; sensitivity to CSD temperature profiles \citep{Li.2022}; and distinct accretion variability \citep{Lai.2022}. 
Moreover, a tightly-separated embedded binary still accretes from a large 3D structure (see Section \ref{sec:ang} below) whereas an isolated binary is mostly accreting in a 2D manner -- unless it is inclined \citep{Moody.2019}. }

\subsection{The Accretion scale} \label{sec:accretion_scale}

Finally, at the smallest scales there is the accretion surface near $\sim {\rm few} \, r_g$ of each BH. 
Material arrives here by either accreting through the large CSD or by low-angular momentum flows sourced from higher altitudes in the AGN disk. 
If we assume that the BH captures as little as $1\%$ \rev{of the material accreting across its orbit, and that the SMBH accretes at $1\%$ of the Eddington rate, then the resulting BH accretion rate may be $\gg 10^4$ times the BH's Eddington rate}\footnote{\rev{If there is a population of accreting objects in the AGN disk, we would expect a radially increasing profile of the disk's inwards radial accretion rate. Thus, the disk accretion rate at our target BH's location is likely to be larger than the value the SMBH accretes at.}}.  
Such a super-Eddington flow will surely generate a strong outflow on the horizon scales. 
This may be in the form of a jet, or winds from the surface layers of the CSD. 
In either case, there is an expectation that this feedback will propagate back up the length scale hierarchy -- possibly all the way to the AGN disk scales. 
If this feedback is strong enough, it may alter the flow on the scale of the Hill sphere and provide a way for the BH to self-regulate its accretion rate to be more Eddington. 

From this it is clear that the problem of BBH-disk interaction involves all length scales from the horizon scale to the global AGN scale. 
In this work, however, we focus on the larger scale problem and in particular we focus on understanding how one parameter, $a_b/R_H$, determines both the orbital evolution rate and accretion rate of the BBH. 
Using Figure \ref{fig:agn} as a guide, we limit ourselves to studying binaries with $R_H/H=0.8$ and $a_b/R_H \sim 0.1 - 0.3$. 
These numbers are plausible for an e.g., $60 M_\odot$ BBH in orbit around a $10^8 M_\odot$ SMBH at $\sim 10^4 R_g$. 
For these values it is computationally feasible to evolve the AGN disk and the CSDs around each BH, and as previously mentioned, resolving the CSDs is critical to obtaining sensible gravitational torques \citep{Munoz.2019,Moody.2019,Duffell.2020,Dittmann.2021,Li.2021}.

\section{Numerical Method} \label{sec:method}

In this Section we describe 3D hydrodynamical simulations of embedded BBHs with the publicly available code Athena++ \citep{Stone.2020}. 
Because we focus on BBHs with $R_H < H$, we make use of the shearing-box approximation \citep{Hawley.1995,Stone.1996} that expands the equations of motion about a Cartesian patch co-rotating with the BBH's center-of-mass (COM).
The background shear of the disk is taken into account to approximate the flow of the global AGN disk. 

In this work, we assume the gas is isothermal and do not include an explicit viscosity.
These simplifications are useful to limit the range of physical processes included in what are the first 3D shearing-box simulations of embedded BBHs in AGN disks\footnote{It should be noted that the 3D simulations of \citet{Kaaz.2021crx} only include a shearing {\it wind}, which is different from the full shearing-box equations.}.
Future work will expand on this simple model and include additional processes to understand their effect on BBH evolution.

The isothermal shearing box equations describing the time evolution of the gas velocity, $\vec{v}$, and gas density, $\rho$, are,
\be
\frac{D\vec{v}}{D t} + 2\Omega_0 \uvec{z}\times \vec{v}- 3 \Omega_0^2 x \uvec{x} + \Omega_0^2 z \uvec{z }&=& - \frac{\del P}{\rho} - \del \Phi_g ,\\ 
\frac{D \ln \rho}{Dt} &=& - \del \cdot \vec{v} ,
\ee
where $D/Dt = \partial_t + \vec{v} \cdot \del$ is the convective derivative, $\Omega_0$ is the rotation rate of the frame, and $\Phi_g$ is the gravitational potential from the BHs. 
We assume the gas is isothermal so that the pressure $P= c_s^2 \rho = H_0^2 \Omega_0^2 \rho$, with a constant scale height $H_0$ and constant sound speed $c_s$.  
We include the vertical stratification of the disk with the $\Omega_0^2 z \uvec{z}$ term. 
Consistent with the models shown in Figure \ref{fig:agn}, we assume that the local disk is low mass compared to the BBH, and so we neglect both the disk self-gravity and the force of the disk on each BH.

We transform the shearing box equations into a dimensionless form by choosing a time scale equal to $\Omega_0^{-1}$ and a length scale equal to $R_H$. 
Because we do not consider the disk self-gravity or the back-reaction of the gas onto the BBH, the mass scale is irrelevant. 
We are thus free to rescale the density to a convenient value, e.g., so that the total mass in the Hill sphere is equal to the BBH mass. 
With these choices the momentum equation becomes, 
\be
\frac{D\vec{v}}{D t} + 2 \uvec{z}\times \vec{v} - 3 x \uvec{x} +  z \uvec{z } = &-& \rev{ \left( \frac{H}{R_H}  \right)^2 }\del \ln \rho  \\ 
&+& \sum_{i=1}^2 \frac{3 m_i}{m_b} \frac{ (\vec{x}-\vec{x}_i)}{|\vec{x}-\vec{x}_i|^3} , \nonumber
\ee
where now all quantities are dimensionless.
The factor of 3 in the second line comes from replacing $\frac{m_b}{M_{\rm smbh}} \left(\frac{a_0}{R_H}\right)^3 \rightarrow 3$ using the definition of $R_H$.
From this, it is clear that there are two controlling parameters. 
The first is the ratio $R_H/H$ which measures how strong the BBH gravitational forces are compared to the gas pressure, and the second is the size of the BBH orbit, $a_b$, compared to $R_H$. 

We model the potential from each BH using a spline function that is exactly Keplerian outside of a distance $r_s$ to each BH \citep{Springel.2001}. 
Gas that comes closer than $r_s$ both feels a softened potential and is subject to mass removal. 
Our mass removal algorithm closely follows the torque-free method of \citet{Dempsey.2020} which we have extended to 3D and discuss in more detail in Appendix \ref{sec:app_acc}. 
\rev{We use a torque-free sink because our sink radius is many orders of magnitude larger than the true accretion radius of the BH. Thus, any angular momentum accreted onto the BH from the sink radius will be much larger than the true value.}
In \citet{Li.2021}, it was found that $r_s$ needs to be small compared to $a_b$ in order for spirals to form in the CSDs. 
That paper used a softening length of $r_s\approx 0.08 a_b$, but with a Plummer softened potential $\Phi_{\rm bh}^{-1}\propto \sqrt{|\vec{r}-\vec{r}_{\rm bh}|^2 + r_s^2}$. 
The equivalent $r_s$ for a spline softened potential is $\approx 2.8$ times larger than the Plummer softening \citep{Springel.2001}. 
Thus, we adopt a larger softening of $r_s = 0.12 a_b$, and set the mass removal length to be equal to $r_s$ in the BBH orbital plane.

\subsection{Mesh refinement, Initial and Boundary Conditions} \label{sec:ic}

We utilize the (static) mesh refinement capabilities of Athena++ to place high resolution regions around the BBH.
Our domain contains a root grid with resolution $\approx H/2$ that spans $x=[-24 H, 24H]$, $y=[-24 H, 24 H]$, and $z=[-4 H, 4 H]$. 
On top of this root grid, we place four refined regions, each of which has a resolution four times higher than the previous region and are located at $|x,y|<5 R_H$, $|x,y|<2.5 R_H$, $|x,y|<R_H$, and $|x,y|<2.5 a_b$ in the x and y directions.
The cutoffs in the z direction are chosen to provide nearly cubic cells throughout the domain. 
The highest level of refinement has approximately $100$ points per $a_b$ on a side. 
\rev{Because Athena++ always enforces no more than one level of refinement difference between neighboring cells, there are additional, automatically placed, refinement regions in between those specified that provide a gradual transition in resolution.}

We start each simulation with the equilibrium solution, $v_y = -3/2 x$, $v_x=v_z=0$, and $\rho = \rho_0 \exp(-(z^2/2) (R_H/H)^2)$. 
Because we neglect disk self-gravity and the disk-induced binary motion, the choice of $\rho_0$ does not change our results. 
We therefore scale all of our simulations so that the total mass in the BBH Hill sphere is equal to the BBH mass. 
The BBH is gradually introduced to the system by growing its mass from zero to its full value in a time of $\Omega_0 t_{\rm grow} = 0.5$. 
Our boundary conditions are shear periodic in the x direction, and periodic the y- and z-directions.  
We have found that periodicity in the z-direction is better behaved than an outflow boundary condition. 

\rev{Because our shearing boxes are closed and we allow the BHs to accrete, the total mass in the domain decreases over time at a rate equal to the binary accretion rate, $\dot{m}_b$. 
This is not a concern, however, as we will show in Section \ref{sec:accretion}, $m_{\rm tot}/\dot{m}_b \sim 3500 \Omega_0^{-1}$ to within a factor of a few, where $m_{\rm tot}$ is the total initial mass in the domain ($m_{\rm tot} \sim 750 m_b$). This time scale is much longer than the time for the dynamics in the Hill sphere to reach a quasi-steady-state, which we have found to be no more than $20 \Omega_0^{-1}$. 
}

\subsection{N-Body Integration} \label{sec:nbody}

\begin{figure}
\centering
\includegraphics[width=.48\textwidth]{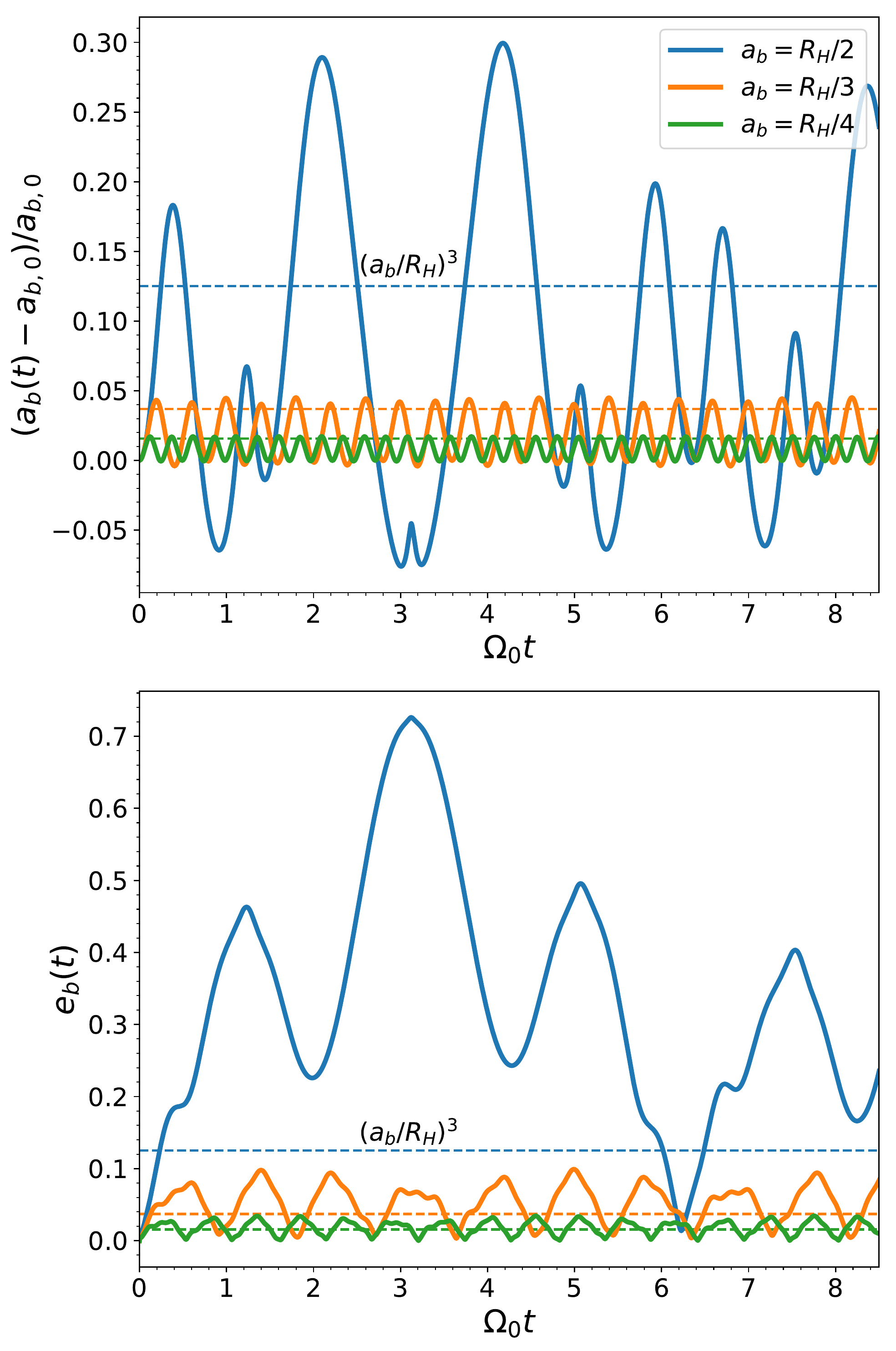}
\caption{Time evolution of $a_b$ (top) and $e_b$ for three example BBH-SMBH systems with initial binary separations of $a_b/R_H = 1/2,1/3$, and $1/4$. 
The tidal force from the SMBH induces an eccentricity and causes periodic oscillations in $a_b$ and $e_b$. 
The dashed horizontal lines mark the estimated magnitude ($(R_H/a_b)^3$) of the $a_b$ deviations and eccentricity. 
The BBH with separation $a_B = R_H/2$ experiences very high eccentricity due to the triple instability \citep{Eggleton.1995,Mardling.2001,Li.2021}. 
The eccentricity in this case rises to a level where our simple estimation is no longer valid. 
The more tightly separated binaries maintain the expected eccentricity. 
}
\label{fig:nbody}
\end{figure}

During a simulation, we evolve the binary separation $\vec{r}_b = \vec{r}_2 - \vec{r}_1$ and binary velocity $\dot{\vec{r}}_b$ with a simple drift-kick-drift scheme. 
We choose this simple integrator over more complicated integrators for two reasons.
First, the hydrodynamical time step is very small compared to the BBH orbital period. 
Typical values are in the range of $\Omega_0 \Delta t \approx 10^{-6} - 10^{-5}$ which amount to at least tens of thousands of steps per orbit. 
With such a small time step, the errors in a low order symplectic integrator will be negligible. 
Second, a drift-kick-drift scheme naturally couples with the Van-Leer second order integrator used in Athena++ \citep{Stone.2020}. 
The Van-Leer integrator consists of two stages: a prediction step where the conserved variables are advanced from the beginning of the time step ($\vec{U}^n$) to the half time step ($\vec{U}^{n+1/2}$) according to, 
\be
\vec{U}^{n+1/2} = \vec{U}^n + \frac{\Delta t}{2} \vec{F}(\vec{U}^n) ,
\ee
and a correction step that advances $\vec{U}^n$ to the end of the time step ($\vec{U}^{n+1}$) following,
\be
\vec{U}^{n+1} = \vec{U}^n + \Delta t \vec{F}(\vec{U}^{n+1/2}) .
\ee
During the prediction step of the hydro integrator, the BBH is ``drifted" to its position at the half time step (analogous to determining $\vec{U}^{n+1/2}$). 
And then during the correction step, the BBH is ``kicked" to its new velocity at the full time step using the positions at the half time step (analogous to evaluating $\vec{F}(\vec{U}^{n+1/2})$).
During the kick step, we add the gravitational accelerations between the disk and each BH to a running time average of $\dot{a}_b$. 
Note that we do not keep track of the kicks in the prediction step, since only the kicks in the correction step actually change the fluid momenta. 

The equation of motion for the binary separation is, 
\be \label{eq:bbh_com}
\ddot{\vec{r}}_b = - \frac{G m_b}{r_b^3} \vec{r}_b + \delta \vec{a}_{\rm smbh} + \delta \vec{a}_{\rm disk} .
\ee  
The first term is the Keplerian acceleration of the binary, while the last two terms are the tidal accelerations due to the gravitational interactions with the central SMBH and the disk. 
All of our simulations include the SMBH tidal acceleration which can be written as,
\be
\delta \vec{a}_{\rm smbh} = G M_{\rm smbh} \left[ \frac{\vec{R} - \vec{r}_2}{|\vec{R}-\vec{r}_2|^3} - \frac{\vec{R}-\vec{r}_1}{|\vec{R}-\vec{r}_1|^3} \right] ,
\ee 
where $\vec{r}_1 = -\mu_2 \vec{r}_b$, $\vec{r}_2 = \mu_1 \vec{r}_b$ and $\mu_{1,2}= m_{1,2}/m_b$.
The separation $\vec{R} = \vec{r}_{\rm smbh} - \vec{r}_{\rm com}$ denotes the separation of the outer SMBH-COM binary. 
\rev{To keep track of the phase of the outer binary, we have two options. 
One is to evolve the inner binary in the rotating frame of the outer binary by adding the appropriate rotating frame forces to Equation \eqref{eq:bbh_com}. 
The other is to stay in the non-rotating frame and solve the equation of motion for $\vec{R}$ which is given by, }
\be \label{eq:agn_com}
\ddot{\vec{R}} = - G (M_{\rm smbh} + m_b) \left[  \mu_1 \frac{\vec{R}-\vec{r}_1}{|\vec{R}-\vec{r}_1|^3} + \mu_2 \frac{\vec{R}-\vec{r}_2}{|\vec{R}-\vec{r}_2|^3} \right] .
\ee
\rev{Both methods are equivalent. For our simulations, we choose to evolve Equations \eqref{eq:bbh_com} and \eqref{eq:agn_com} together. }
When evolving Equation \eqref{eq:bbh_com} we do not include the disk acceleration $\delta \vec{a}_{\rm disk}$. 
However, we still measure $\delta \vec{a}_{\rm disk}$ every time step and use it to measure the orbital evolution rates presented in Section \ref{sec:adot}.

As a test of our integrator, we show in Figure \ref{fig:nbody} the time evolution of the osculating $a_b$ and $e_b$ with initial $a_b/R_H = 1/4, 1/3,$ and $1/2$. 
The non-zero $\delta \vec{a}_{\rm smbh}$ induces oscillations in the BBH orbital energy and a non-zero eccentricity that scale with the initial binary separation.
Approximating $|\delta \vec{a}_{\rm smbh}| \approx 2 a_b/a_{\rm com}^3$, the magnitude of the total BBH acceleration is roughly $G m_b/a_b^2 ( 1 - (2/(G m_b)) (a_b/a_{\rm com})^3)$. 
We find that the fractional deviation of $a_b$, i.e. the eccentricity, induced by the AGN is roughly $\Delta a_b/a_b \sim (a_b/R_H)^3$. 
For $a_b/R_H = 1/4$, and $1/6$, we would expect eccentricities and deviations $\Delta a_b/a_b$ of $\sim 0.005$ and $0.015$, respectively. 
These estimates are in good agreement with the numerical results for $a_b/R_H < 0.5$. 
When $a_b/R_H \gtrsim 0.5$, however, $e_b$ reaches very large values due to the known triple instability \citep{Eggleton.1995,Mardling.2001,Li.2021}.
Because of this we focus only on binaries with $a_b \le R_H/3$.

In all of our simulations, we fix the SMBH-COM binary orbit to be circular with mean motion $\Omega_0$. 
One could in principle extend our results to non-circular SMBH-COM binaries by working in the guiding center frame where the BBH COM executes an orbit about the center of the shearing box. 
But, the SMBH-COM eccentricity would have to be small (e.g., $<H$) in order to be consistent with the shearing-box assumptions. 
In this paper we focus only on circular, equal-mass BBHs. 
In future works we will relax these assumptions to examine non-equal mass ratio, eccentric, and inclined BBHs.

\subsection{Tracking Angle Averaged Profiles} \label{sec:profiles}

In Section \ref{sec:results}, we present time and azimuthally averaged profiles of mass accretion, momentum, and torque density centered on each BH.
This is a non-trivial task, as these profiles are computed {\it in-situ} around continuously moving objects.
Here we outline our procedure.

Our goal is to measure the angle- and time-averaged profile of a quantity $\avg{\rho q}$ , where $q$ can be a velocity, a specific torque, etc.
To do this, we attach a Lagrangian grid to each BH. 
This grid is either a 2D cylindrical $(R,z)$ grid, or a 1D spherical grid. 
Each location in either of these grids can be thought of as a bin in which we will collect the total amount of $\rho q$. 
Each cell in the simulation is treated as a particle with mass $\delta m = \rho \delta x \delta y\delta z$ and during every time step we add $\Delta t q \delta m$ to any bin that overlaps with a cell. 
With each simulation snapshot, we dump the time-averaged profiles of each quantity to a file, and then post-process each profile by creating a cumulative sum and then differentiating that sum on a coarser grid. 
In the end, we are left with time and angle-averaged profiles of e.g., density, specific angular momentum, mass flux, or torque density using every time step of the simulation. 
In future work, we will also use these Lagrangian grids to incorporate feedback from each BH.

\section{Results} \label{sec:results}

In this Section, we present the results from 3D embedded BBH simulations at five different separations, $a_b/R_H = 1/3$, $1/4$, $1/6$, $1/8$, and $1/10$.
For each simulation, we initialize the binaries on circular orbits with mean motions $n_b/\Omega_0 = 9, 13.9, 25.5, 39.2$, and $54.8$, respectively, where $n_b = \sqrt{G m_b/a_b^3}$. 
Each simulation is run until the gas in the vicinity of the BBH is in steady-state. 
Our measure of steady-state is that the time and angle averaged radial mass flux (centered on each BH) is spatially constant with a value equal to the BHs' accretion rates. 
This typically occurs within a time of $\Omega_0 t \sim 6$.
\rev{In order of decreasing $a_b$, the simulations are run for total times of $\Omega_0 t_f = 20, 20, 20, 12.5$, and $6.2$, and final time-averages of quantities are calculated over the last $\Omega_0 \Delta t= 4.5,4.5,4.5, 2.5$, and $0.5$.}
We discuss this convergence in Section \ref{sec:accretion} below.

\begin{figure*}
\centering
\includegraphics[width=.98\textwidth]{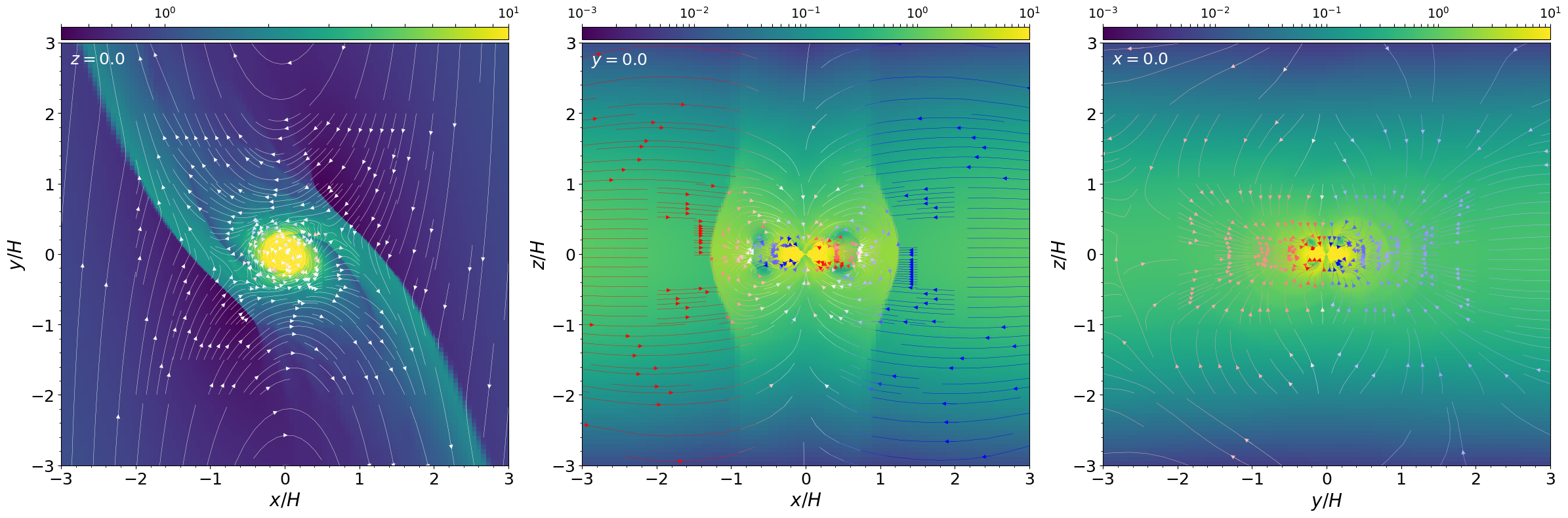} \\ 
\includegraphics[width=.98\textwidth]{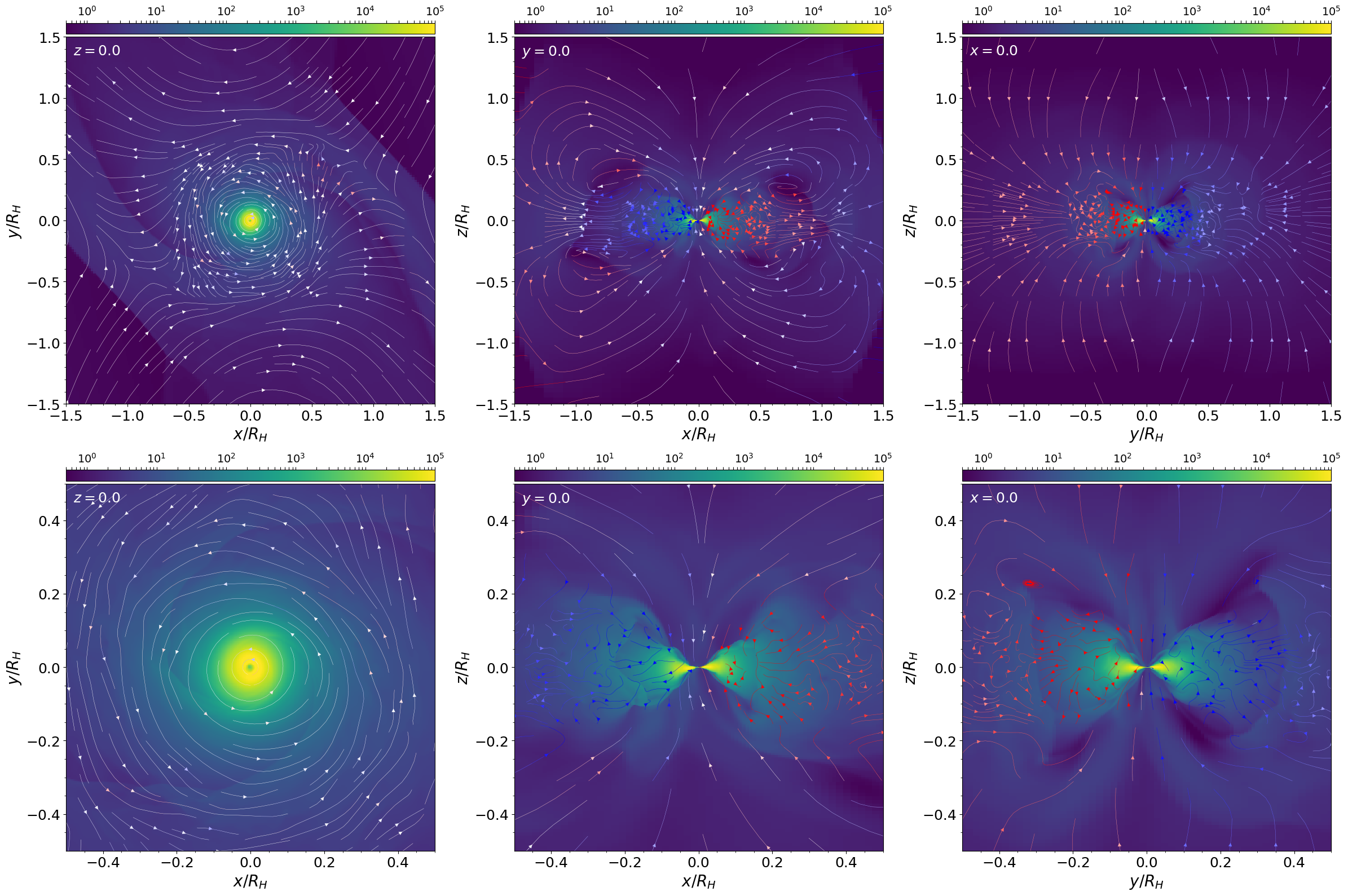}\\
\caption{Snapshots of the gas density and stream lines around a single BH. Each streamline is colored by the out-of-plane velocity with red denoting positive and blue denoting negative values. Each row plots three slices at increasingly smaller distances to the BH. There are inner and outer spiral arms on scales of $H$ and a rotating disk around the BH of thickness of $\sim 0.2 R_H$ on scales of $\sim R_H$. There is a faint $m=2$ spiral in this disk that extends all the way down to the sink radius.
\rev{Additionally, the streamlines in the left column show that the BH's disk is rotating in a prograde sense with its angular momentum vector pointing along the positive $z$-axis.}
 }

\label{fig:gas_single}
\end{figure*}
\begin{figure*}
\centering
\includegraphics[width=.98\textwidth]{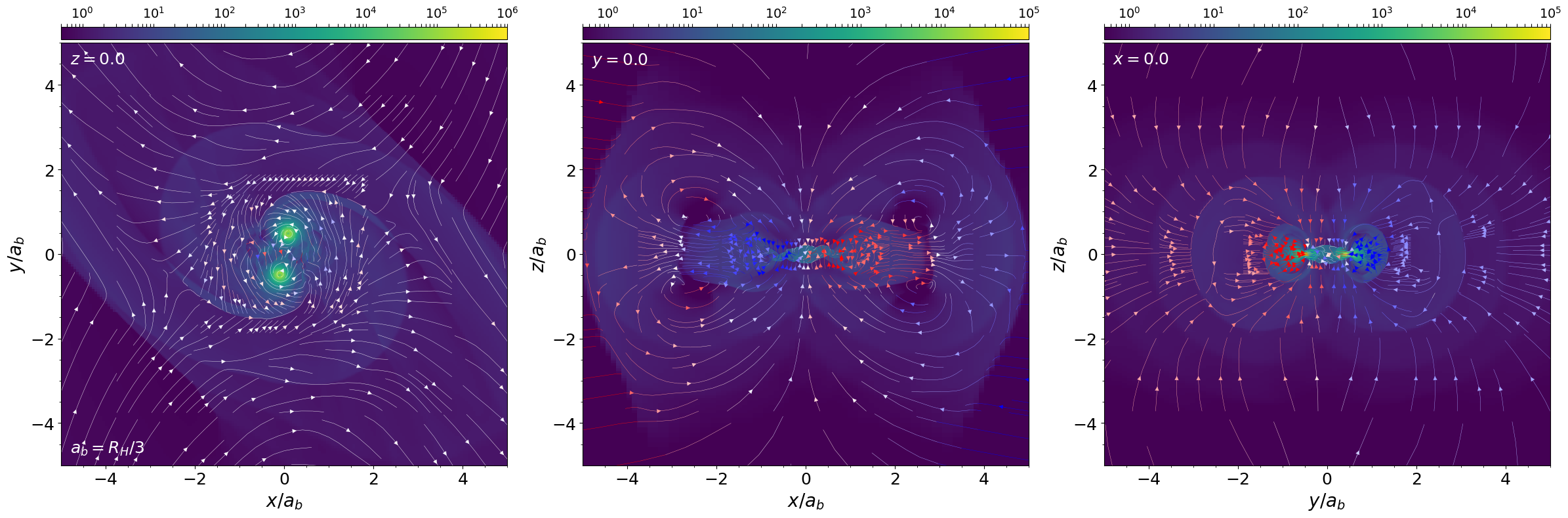}
\includegraphics[width=.98\textwidth]{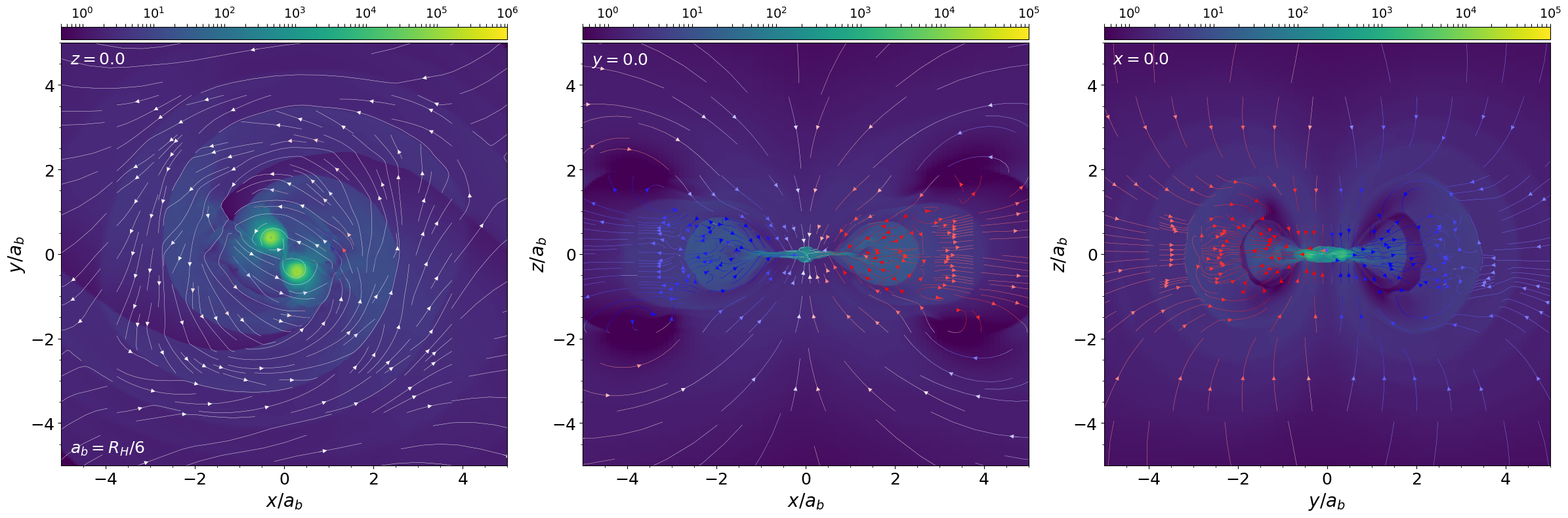}
\includegraphics[width=.98\textwidth]{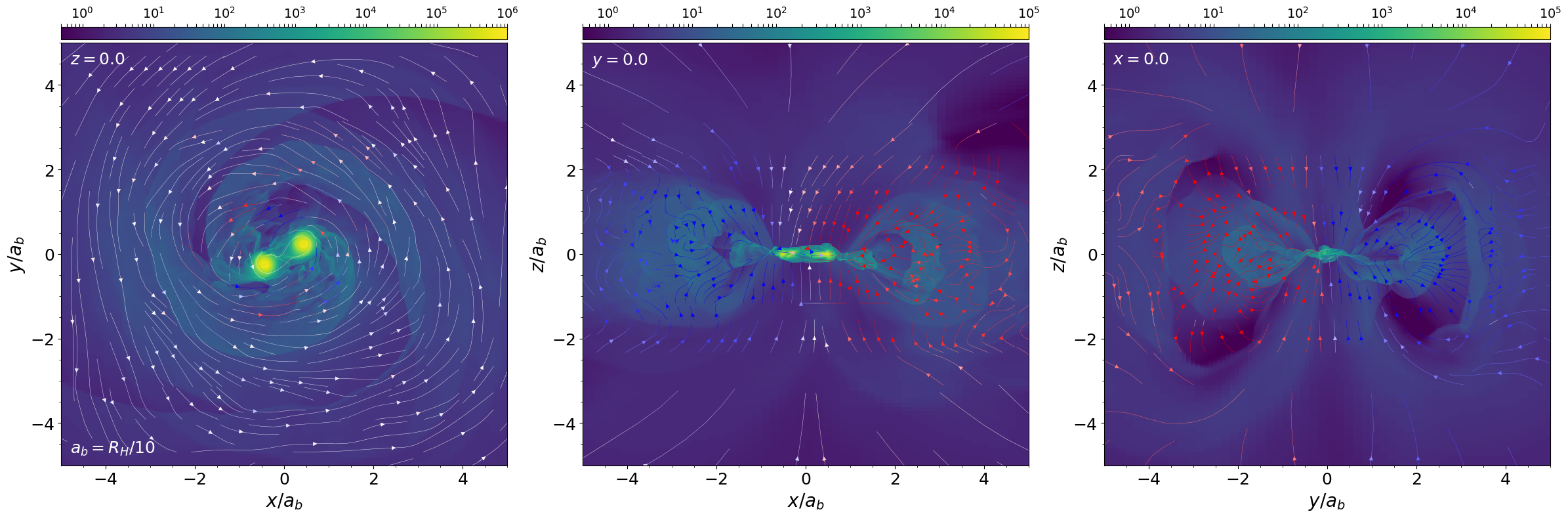}
\caption{Snapshots of the density and stream lines around BBHs with $a_b/R_H=1/3, 1/6$, and $1/10$. 
There is a broad similarity between the three simulations. 
Each has an $m=2$ spiral connecting two \rev{prograde} CSDs to a \rev{pgrograde} CBD, each of which have a thickness of roughly $\sim a_b$ . 
The main difference is the larger CBD around the tighter binaries.
}
\label{fig:gas_bbh}
\end{figure*}

\subsection{Gas morphology}  \label{sec:gas}

To separate out the flow characteristics associated with the BBH COM from the flow induced by the binary motion, we show several slices of the gas density and stream lines from a simulation with one BH\footnote{A single BH can also be thought of as a BBH with $a_b \ll r_s$, i.e a BBH with separation on the scale of $r_g$.} in Figure \ref{fig:gas_single} and a BBH in Figure \ref{fig:gas_bbh}. 
The slices show the mid-plane ($z=0$), and two vertical cuts at $x=0$ and $y=0$. 

Figure \ref{fig:gas_single} shows the gas morphology on three scales. 
On the largest scales ($\gg H$), there are clear inner and outer trailing spiral arms that originate from $\sim H$ away from the BH. 
These spirals are tidally excited by the interaction of the BH with the SMBH and are equivalent to the spirals in shearing box simulations of planet-disk interaction \citep[e.g.,][]{Dong.2011ama,Dong.2011}.
On these scales, gas either executes circulating orbits around the SMBH or librating, horseshoe orbits at distances $|x| \lesssim 2 R_H$. 

On scales of $\sim R_H$ and smaller, the flow is strongly perturbed by the BH. 
Gas either crosses the BH orbit through horseshoe turns, or is captured into circulating orbits around the BH. 
This captured gas builds a sizable rotating disk around the BH (which for simplicity we refer to as the CBD). 
The large scale $m=2$ spiral can be faintly seen in the CBD going all the way down to the sink scale. 
 
The thickness of this disk follows from vertical hydrostatic equilibrium where, 
\be
-\frac{1}{\rho} \ppderiv{P}{z} = \frac{G m_b}{r^3} z + \Omega_0^2 z ,
\ee
and where $r$ is the distance to the BH.
Defining the vertical pressure scale height of the CBD as $h \equiv (-d \ln P/dz)^{-1}$, we find a radial profile of,
\be
h(r) = \frac{H_0}{\sqrt{3}} \left(\frac{r}{R_H}\right)^{3/2} \left[1 + \frac{1}{3} \left(\frac{r}{R_H} \right)^3 \right]^{-1/2} ,
\ee
where $H_0 = c_s/\Omega_0$ is the AGN disk scale height. 
The factor in brackets ensures that $h(r) \rightarrow H_0$ when $r\gg R_H$. 
At distances of $\sim 0.2 R_H$, the vertical thickness of the CBD is on the order of $5 \%$ of $H_0$, which by eye is in agreement with Figure \ref{fig:gas_single}. 
Because we do not add additional heating around the BH, the vertical thickness of the CBD can be very small on the mass removal scale.
This can be partially alleviated by enforcing a rising temperature profile around the BH, as is done in \citet{Li.2022} and simulations of isolated binaries \isolatedp{e.g.,}.

Looking at the vertical flow near $R_H$, there are two distinct patterns. 
In the $y=0$ slice, which is along the line connecting the BH and SMBH, there is a clear circulation pattern where material arrives in the mid-plane from higher altitudes and is pushed outwards -- away from the BH -- where it is then lifted up to join the flow falling flow onto the BH. 
The $x=0$ slice, which goes through the horseshoe region, is different. 
Here, the flow is convergent onto the BH in all directions. 
Material both falls onto the BH directly and indirectly through the CBD.

Figure \ref{fig:gas_bbh} shows the same three slices as Figure \ref{fig:gas_single}, but for three BBHs with $a_b/R_H = 1/3,1/6,$ and $1/10$. 
On the largest scales, the flow and density are nearly identical to the single BH simulation with the same mass as the BBH. 
But, inside of the Hill sphere, the gas morphology changes. 
The CBD is now split into two circumsingle disks  orbiting each binary component. 
What little remains of the CBD now has a large $m=2$ spiral connecting each BH to the large scale spirals. 
Looking at the binary edge-on, we find that the CBD is of similar thickness to the CBD in the single BH case. 
The flow within the Hill sphere is also similar to the single BH case. 
The only major difference is that gas falls onto each BH as opposed to the BBH center of mass. 

The structure of the CSDs is similar to the structure of the single BH CBD. 
They are rotating disks with a characteristic thickness governed by vertical pressure balance and built up by vertical accretion. 
Compared to the CBD, these disks are thinner due to both a lower BH mass and because the distance scales are smaller -- a fraction of $a_b$ compared to a fraction of $R_H$ (see also Figures \ref{fig:mdot_2D_prof} and \ref{fig:mdot_zoom} in Sections \ref{sec:accretion}-\ref{sec:ang} below). 
Each CSD contains an $m=2$ spiral that is tidally excited by the companion BH. 
The outer arms of each CSD spiral join with the $m=2$ spiral going through the CSD, while the inner arms connect in the region between the BHs.
We defer a more detailed discussion of the CSDs until Section \ref{sec:ang}.

From these snapshots, it is clear that the disks around the BHs are built up from vertical accretion of gas from the AGN disk. 
In the next Section, we examine this process in more detail and connect the disk accretion profile to the BBH accretion rate.

\subsection{Accretion} \label{sec:accretion}

\begin{figure}
\centering
\includegraphics[width=.48\textwidth]{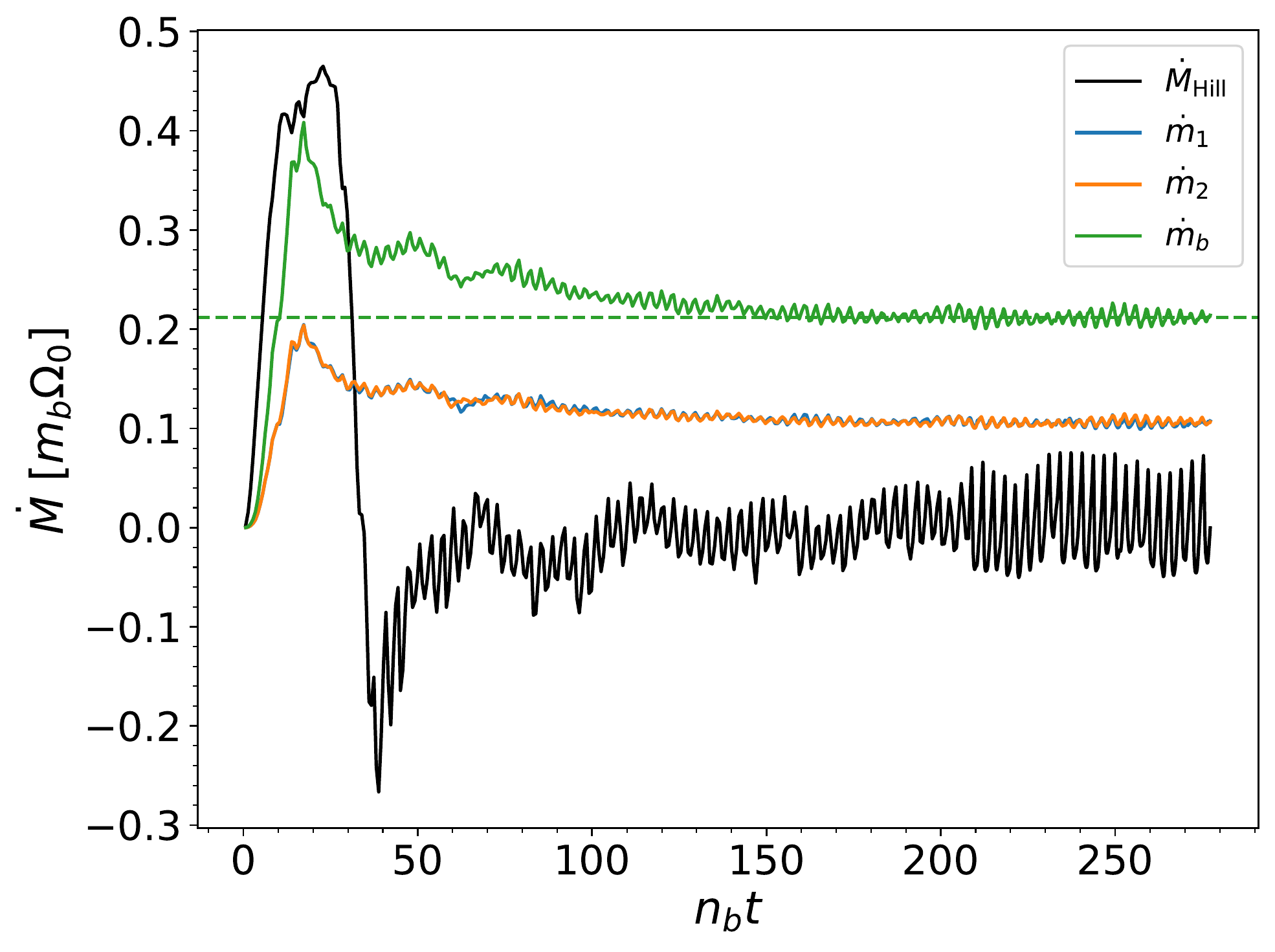}
\caption{Time evolution of $\dot{m}_b$ (blue, orange, and green lines) and \rev{$\dot{M}_{\rm Hill} = dM_{\rm hill}/dt$ (black line), which is the rate at which the mass within the Hill sphere is changing} for the simulation with $a_b = R_H/4$. 
The mass within the Hill sphere reaches a steady-state value in a time of $\sim 100 n_b t$. 
\rev{Not long after,} the $\dot{m}$ onto the BBH reaches a steady-state value.
The individual accretion rates of each BH are equal, as expected for an equal-mass, circular binary.}
\label{fig:mdot_time}
\end{figure}

\begin{figure}
\centering
\includegraphics[width=.48\textwidth]{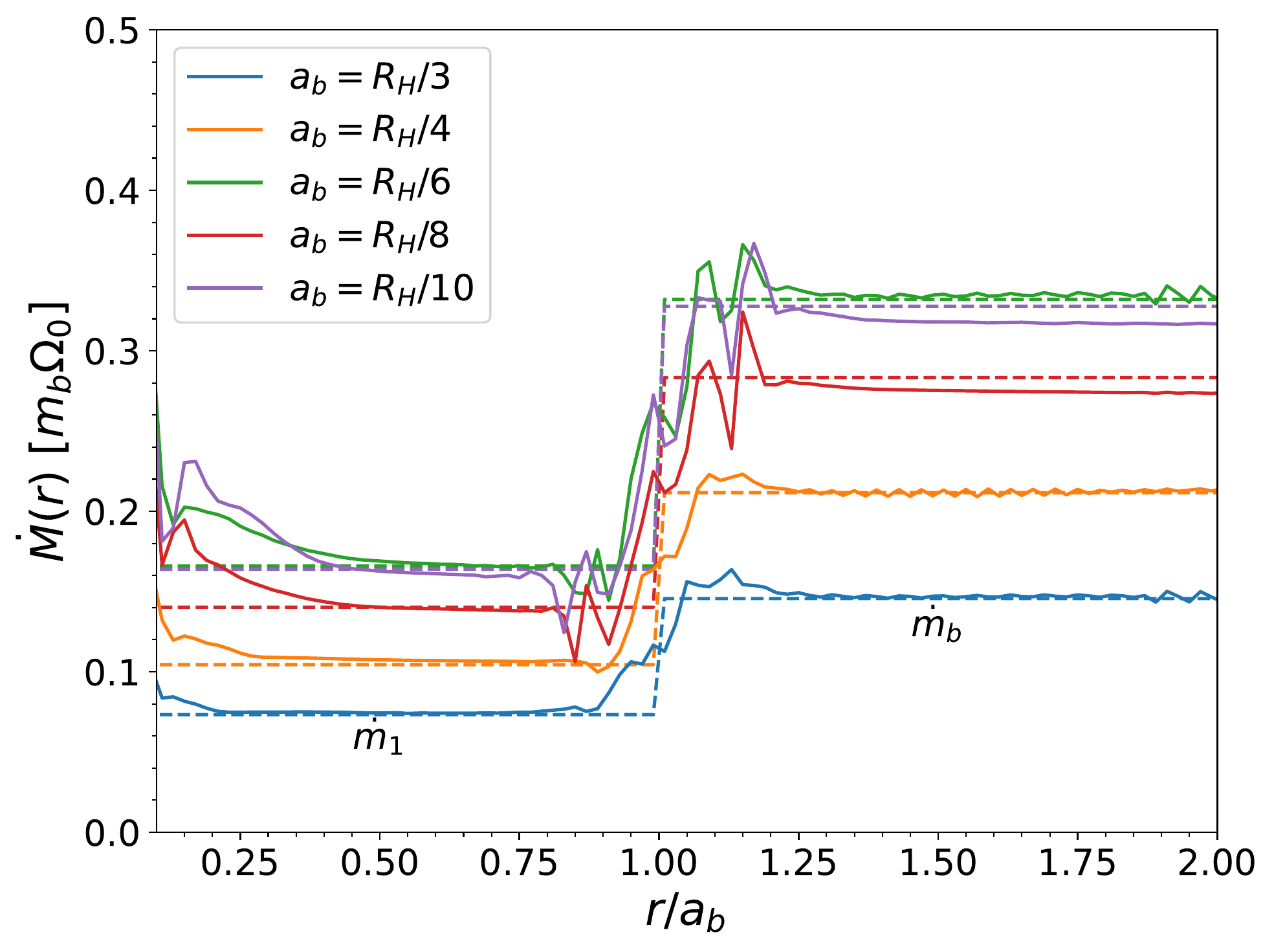}
\caption{Final time- and angle-averaged $\dot{M}(r)$ profiles (solid lines) in the disks around one BH for each of the simulations. The curves are functions of spherical radius from the BH. 
The companion BH is at $r=a_b$.
The dashed lines are equal to the measured $\dot{m}_b$ onto the BBH at $r>a_b$ and the measured $\dot{m}$ onto the primary BH at $r<a_b$. 
The fact that the $\dot{M}(r)$ curves are spatially constant and equal to the measured accretion rates onto the BHs is an indication that the CBD and CSDs are in quasi-steady-state.  }
\label{fig:mdot_profs}
\end{figure}

\begin{table}[t]
\centering
\begin{tabular}{r| c c c c c}
$a_b/R_H$ & $1/3$ & $1/4$ & $1/6$ & $1/8$ & $1/10$ \\ \hline
$\avg{\dot{m}_b}/m_b$ & 0.15 & 0.21 & 0.33 & 0.28 & 0.34   \\
$\avg{\dot{m}_2}/\avg{\dot{m}_1}$ & 0.99 & 1.01 & 1.00 & 1.01 & 0.98
\end{tabular}
\caption{Final steady-state, time-averaged values of the BBH accretion rate (in units of $\Omega_0$), and a measure of the accretion asymmetry.  }
\label{tab:mdot}
\end{table}

\begin{figure*}
\centering
\includegraphics[width=.32\textwidth]{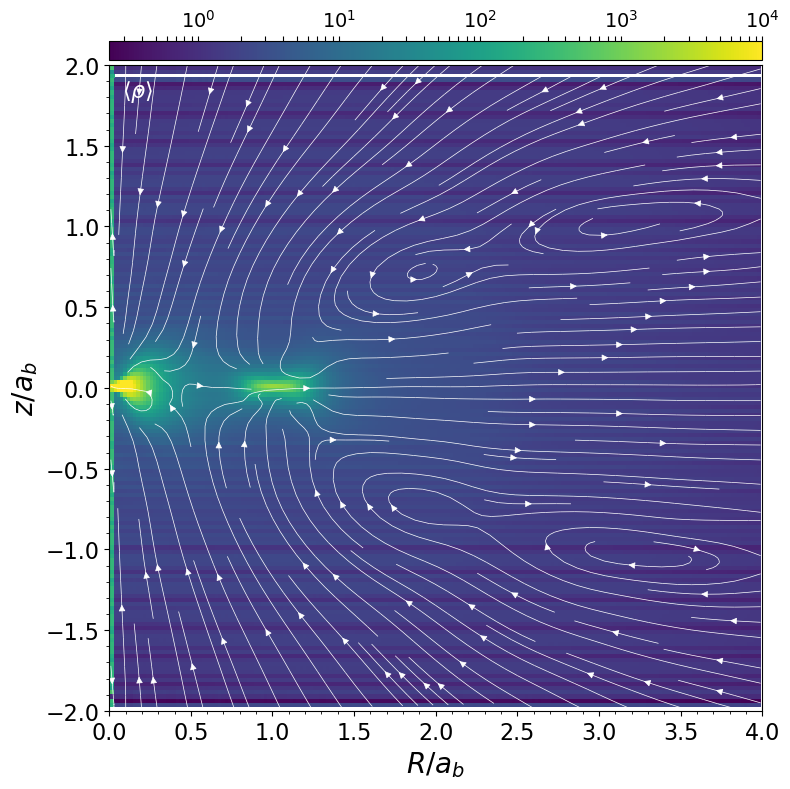}
\includegraphics[width=.32\textwidth]{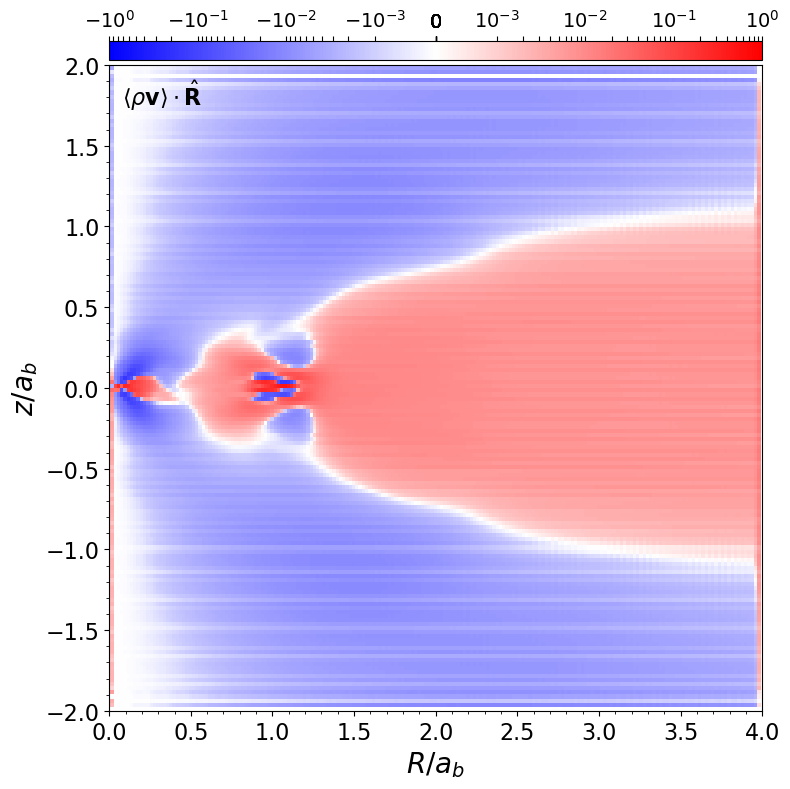}
\includegraphics[width=.32\textwidth]{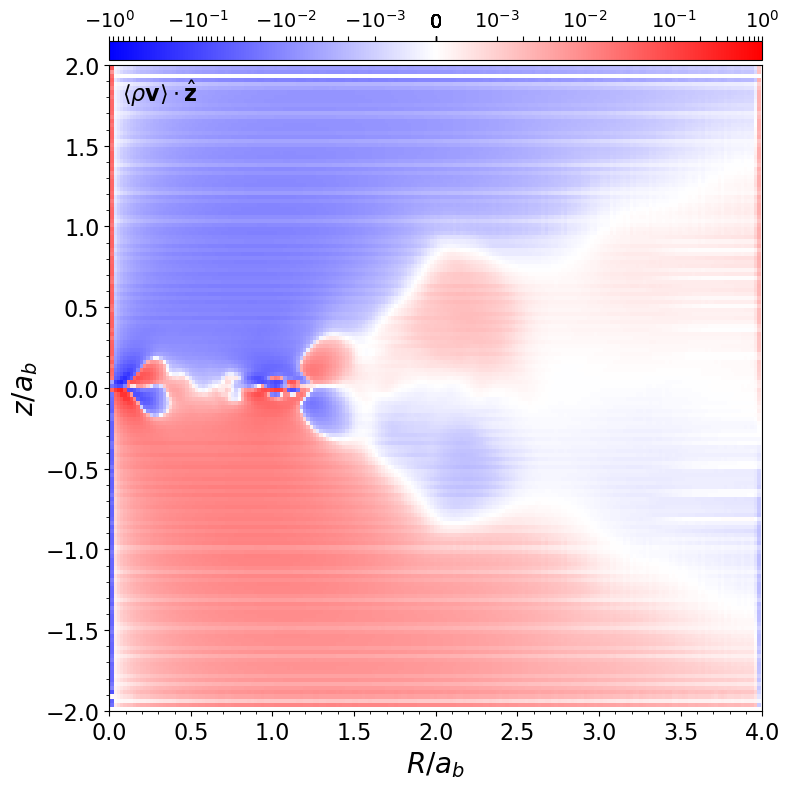} \\ 
\includegraphics[width=.32\textwidth]{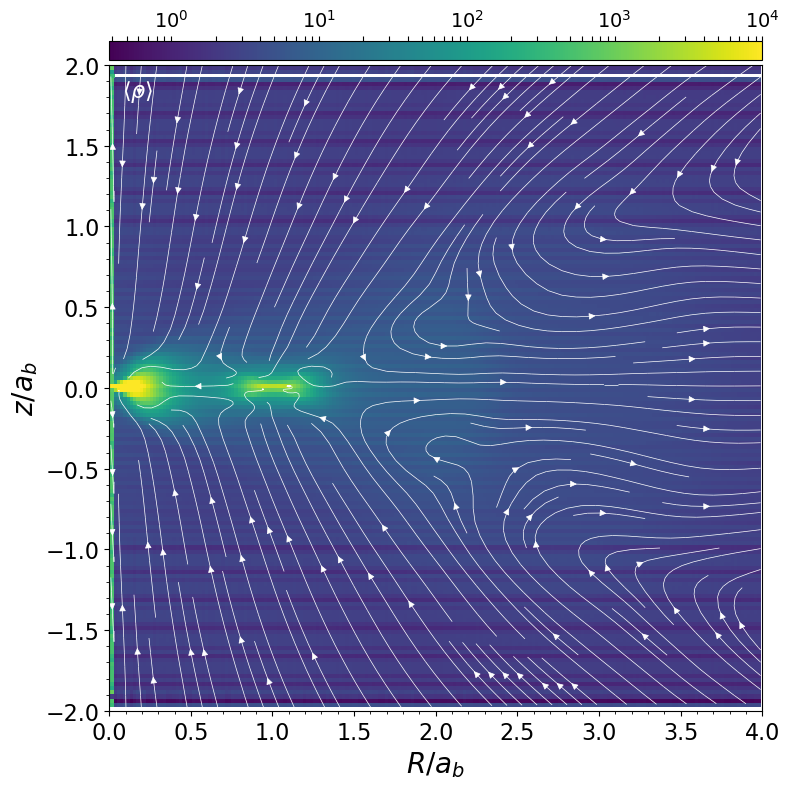}
\includegraphics[width=.32\textwidth]{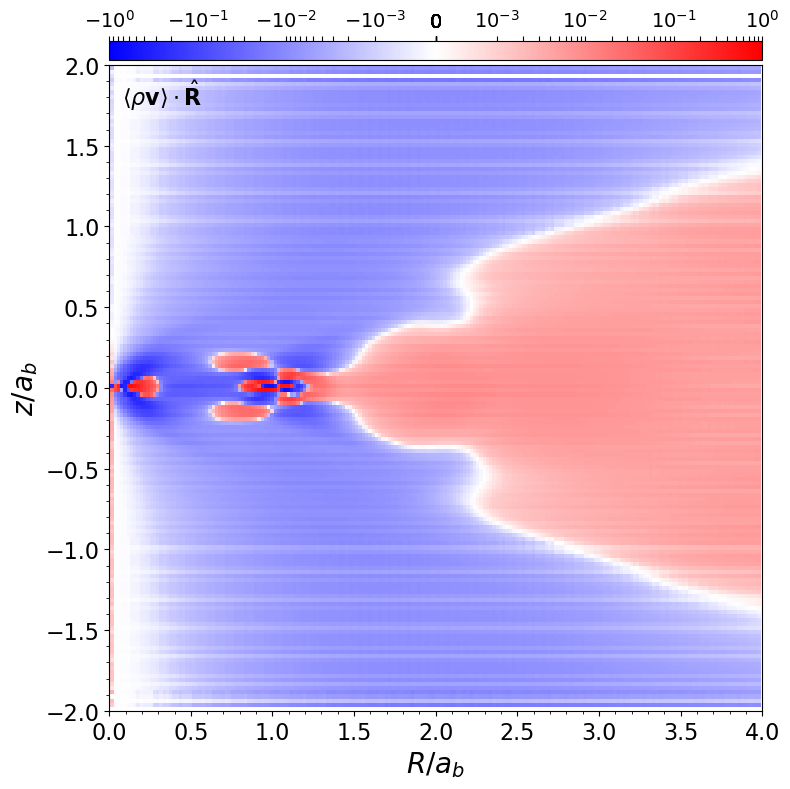}
\includegraphics[width=.32\textwidth]{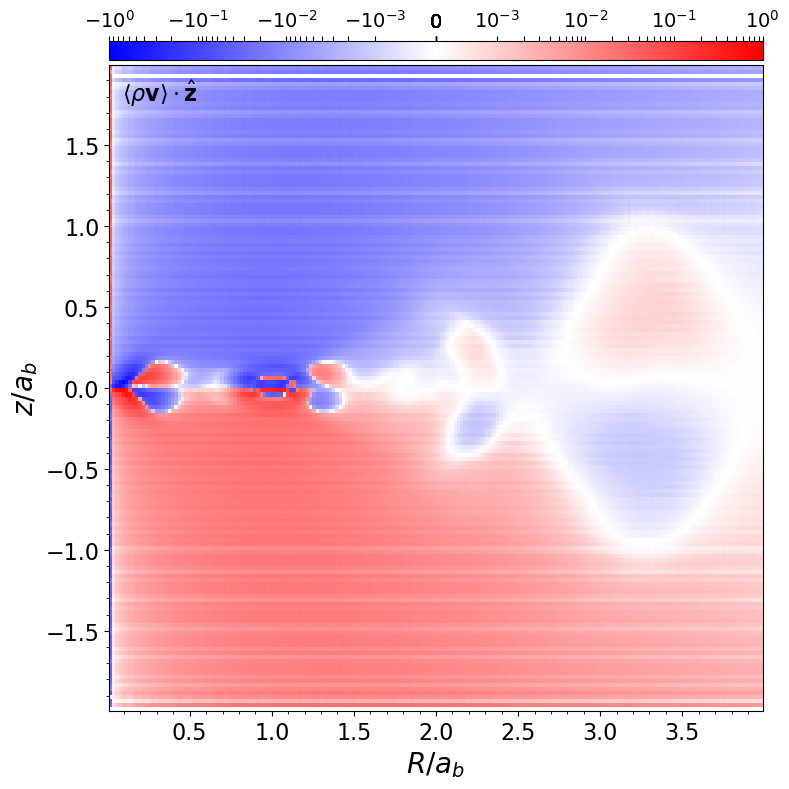} 
\caption{Final time- and azimuthally-averaged $(R,z)$ profiles of density (left), $\dot{\vec{M}} \cdot \uvec{R}$ (middle), and $\dot{\vec{M}} \cdot \uvec{z}$ (right) for BBHs with $a_b=R_H/3$ (top row) and $R_H/6$ (bottom row). 
The stream lines shown on the density plots are computed from the right two columns and indicate how mass is flowing in the vicinity of each BH.
\rev{The horizontal white lines in the $\dot{\vec{M}}$ plots are artifacts of the averaging and plotting process and do not affect any of our conclusions.}
 }
\label{fig:mdot_2D_prof}
\end{figure*}

\begin{figure*}
\centering
\includegraphics[width=.48\textwidth]{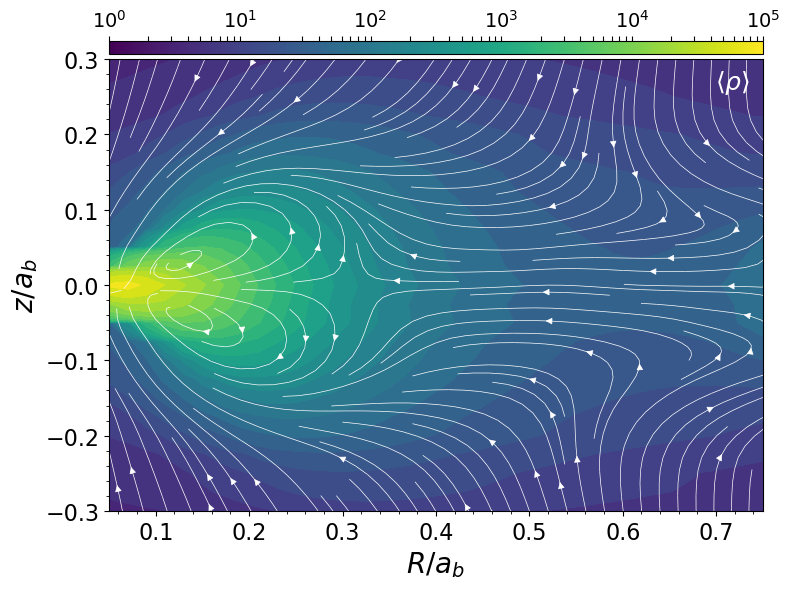}
\includegraphics[width=.48\textwidth]{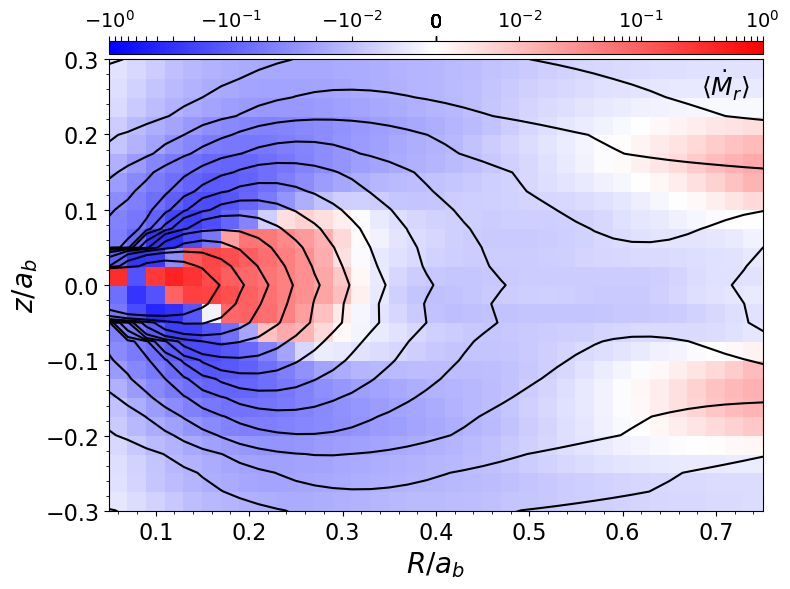}
\caption{Final time- and azimuthally-averaged density (left) and radial $\dot{M}$ (right) profiles around one BH in a binary with $a_b = R_H/4$. The contours shown in the density plot are also shown in the $\dot{M}$ plot as black lines. The isodensity contours are closed indicating that there is a torus around each BH that extends to $\sim 0.45 a_b$. The surface layers of this torus accrete onto the BH, whereas the mid-plane decretes. }
\label{fig:mdot_zoom}
\end{figure*}
\begin{figure}
    \centering
    \includegraphics[width=.48\textwidth]{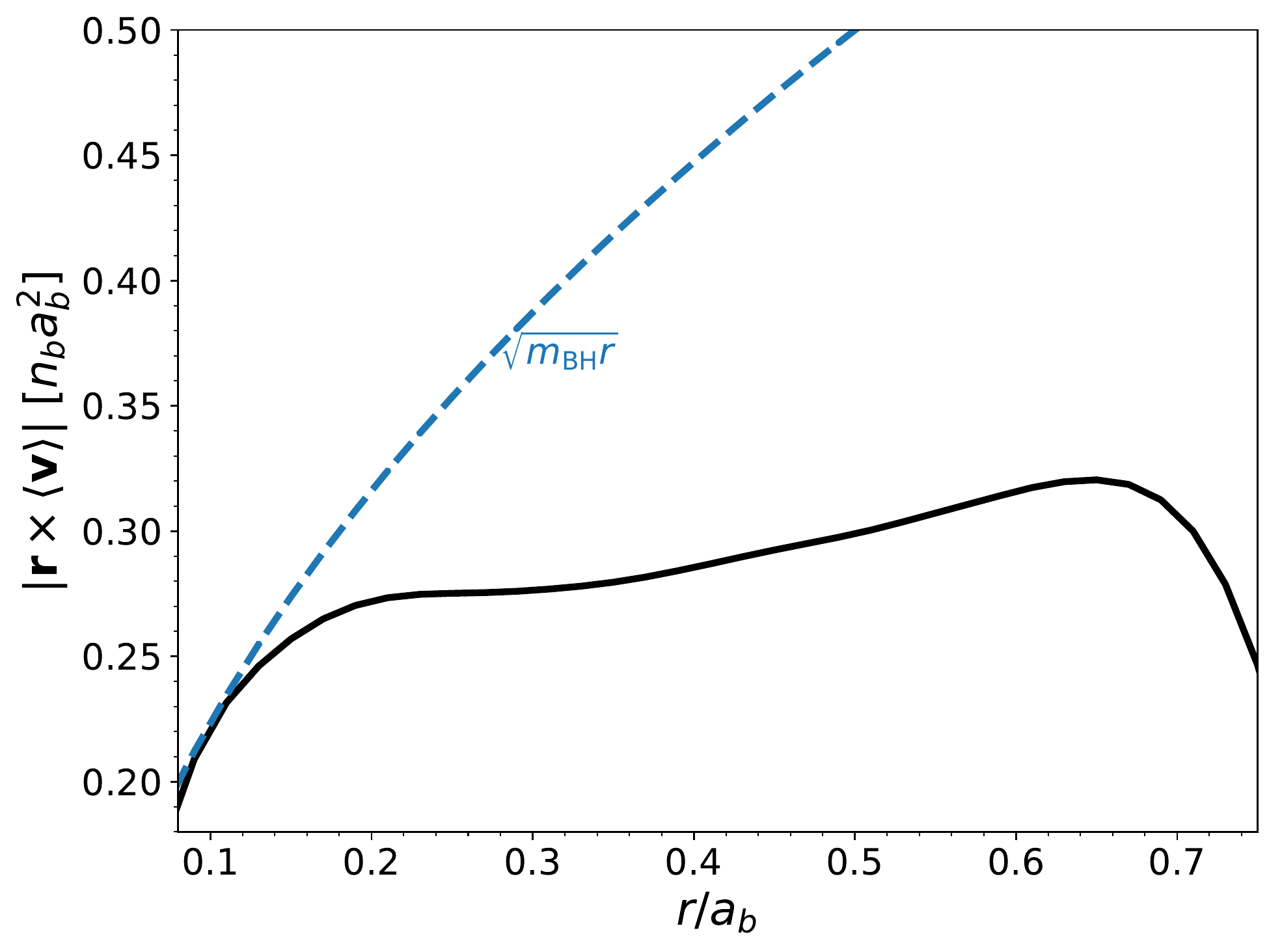} \\
    \includegraphics[width=.38\textwidth]{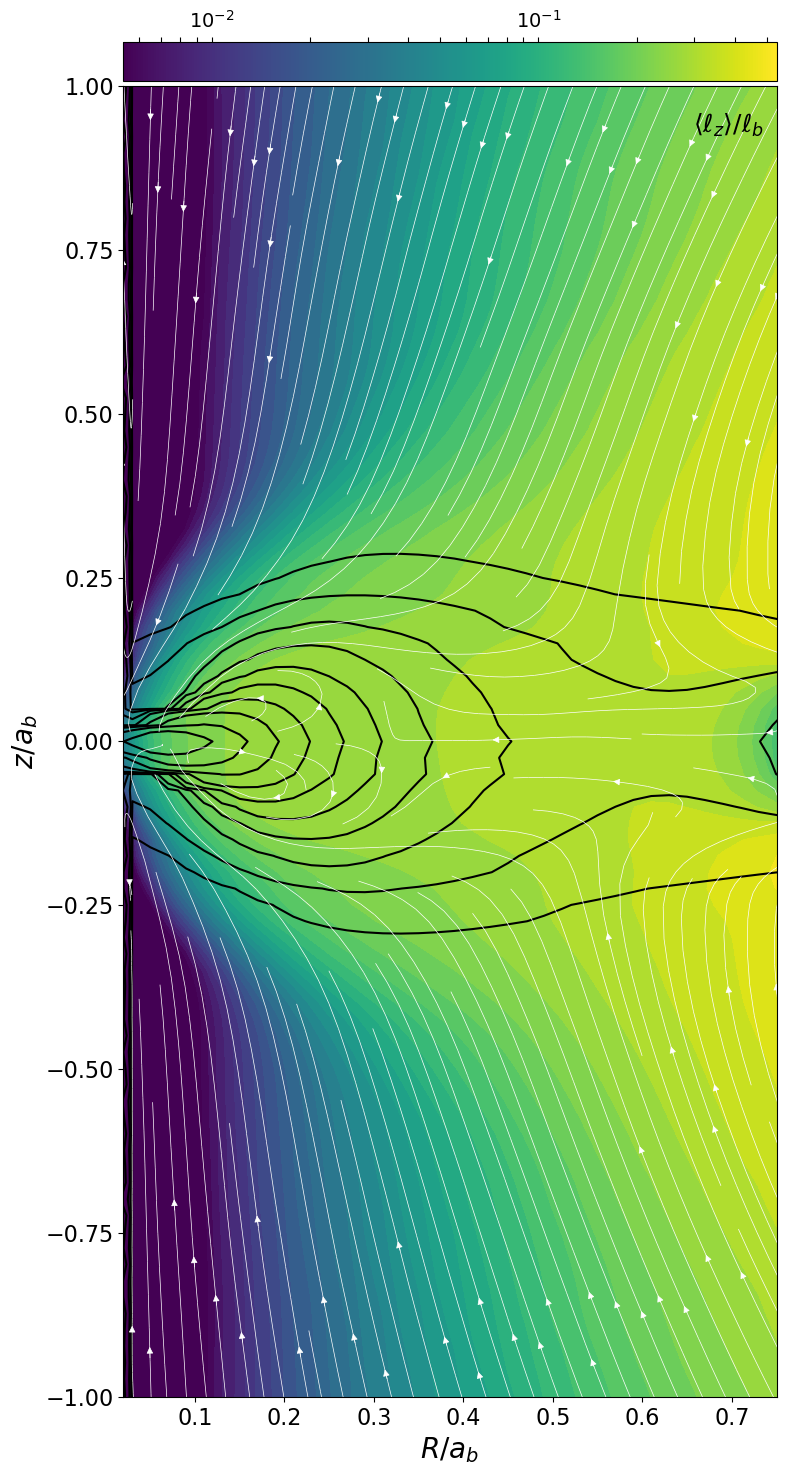}
    \caption{Total specific angular momentum profile as a function of spherical distance to one BH (top) and the vertical specific angular momentum profile, $(\ell_z)$, as a function of $(R,z)$ (bottom) in a binary with $a_b=R_H/4$.
    The blue dashed line in the top panel shows the strictly Keplerian profile.
    In the bottom panel, the stream lines show the flow direction as in Figure \ref{fig:mdot_zoom}, while the black contours show the isodensity lines.
    }
    \label{fig:lprof}
\end{figure}

Accretion of gas onto an embedded BBH is an important process. 
In addition to growing the binary mass, preferential accretion can alter the mass ratio of the binary, and accreted momentum can alter the orbital evolution of the binary.
Moreover, accretion onto compact objects can produce powerful feedback effects such as jets and strong radiation that can alter the AGN disk on a more global scale \citep[e.g.,][]{Jiang.2014,Paschalidis.2021,Wang.2021,Tagawa.2022}. 
Including a prescription for accretion in simulations of embedded BBHs is thus crucial for obtaining sensible results for their growth and orbital evolution.

Figure \ref{fig:mdot_time} shows the evolution of the accretion rate onto each BH as a function of time for one simulation with $a_b = R_H/4$. 
As can be seen in Figure \ref{fig:mdot_time}, an equal mass and circular BBH accretes symmetrically with $\dot{m}_1 \approx \dot{m}_2$, as expected from previous studies \citep[e.g.][]{Munoz.2019,Tiede.2021k6r}.

The accretion rates \rev{effectively} converge in time after $n_b t \sim 150$.
\rev{The total binary accretion rate (shown as the green line)  at this time is equal to its final value (shown as the dashed green line).}
In addition to the BH accretion rates, the mass within the Hill sphere of the BBH \rev{converges after about $n_b t \sim 100$.}
This can be seen from the black curve in the figure which shows the rate at which the total mass in the Hill sphere is changing, \rev{which we denote $\dot{M}_{\rm Hill}$}. 
At early times, the Hill sphere grows at an increasing rate until the losses through accretion onto the BBH slows this growth down and eventually an equilibrium state is reached where $\dot{M}_{\rm Hill} \approx 0$. 
In this steady-state, the $\dot{M}$ filling the Hill sphere is equal to the $\dot{M}$ leaving the simulation through the BHs.

We make this last point more clear by directly comparing the BH accretion rates to the measured $\dot{M}$ through the Hill sphere region in Figure \ref{fig:mdot_profs}. 
Here, the solid curves show the time- and angle-averaged radial mass flux ($\dot{M}(r)$) profiles as a function of the distance to one of the BHs for each of our simulations.
The companion BH sits at $r/a_b=1$ in this plot. 
Exterior to the companion, the $\dot{M}$ profiles are constant and equal to the total $\dot{M}$ onto the binary (shown as the dashed lines). 
Interior to the companion, the accretion rate drops due to the mass lost onto the companion and reaches another roughly constant value that is equal to the measured accretion rate onto the primary BH.
The fact that these $\dot{M}$ profiles are approximately spatially constant and equal to the accretion rates of the BHs is a strong indicator that a steady-state has been reached in the BBH's Hill sphere. 

\rev{At separations smaller than $a_b< R_H/6$, the $\dot{M}(r)$ values inside of $r\lesssim 0.5 a_b$ start to deviate from the values of $\dot{M}(r)$ near $r \sim 0.5 a_b$ by, at most, $40\%$. 
Moreover, for these simulations $\dot{M}(r)$ outside of $r \gtrsim a_b$ is slightly lower than $\dot{m}_b$. 
While these deviations could be an artifact of the spherical averaging process, it is perhaps more likely that for these simulations the smallest scales may not be in perfect steady-state.
Nevertheless, we will show later that the important quantities such as $\dot{a}_b/\dot{m}_b$ do appear to be converged in time.  }

Table \ref{tab:mdot} lists the steady-state values of $\dot{m}_b$ and $\eta = \dot{m}_2/\dot{m}_1$ for each simulation.  
All simulations find that $\eta \approx 1$ to within $1\%$, as expected for equal mass, circular binaries. 
The values of $\avg{\dot{m}_b}$ are within a factor of no more than \rev{ $\sim 2.3$ of each other, ranging between $0.15 m_b$ and $0.34 m_b$. }
This should be somewhat expected if accretion is driven by the flow of mass into the Hill sphere.
So long as the binaries do not strongly perturb this flow, they should all accrete at this rate. 
This is consistent with the largest difference in $\dot{m}_b$ occurring in the widest binaries.
\rev{Moreover, this suggests that our mass removal method is not affecting our $\dot{m}_b$ values.}

To better understand how mass is delivered to each BH, we show the time- and azimuthally-averaged profiles of density and momentum in a cylindrical $(R,z)$ coordinate system centered on one of the BHs in Figure \ref{fig:mdot_2D_prof}\footnote{We will always distinguish the cylindrical radius from the spherical radius by capitalizing the former.}.
We focus on the two simulations with $a_b = R_H/3$, and $R_H/6$. 
The momentum plots show the $\uvec{R}$ and $\uvec{z}$ components. 
In the density plot, the stream lines are constructed from the two momentum plots.
First, on scales larger than $a_b$, material flows away from the BBH in the orbital plane. 
This flow is driven by material falling from above and below that is being turned around at the outskirts of the companion BH's CSD. 
There are additional vortical structures at $z\sim \pm a_b$ and $R\sim R_H$ that were also seen in Figures \ref{fig:gas_single} and \ref{fig:gas_bbh}. 
By and large, mass is being delivered to the binary region from $|z|> a_b$ -- which is similar to what is seen in 3D studies of planet-disk interaction \citep{Fung.2016}. 

\subsection{The Black Hole Torus} \label{sec:ang}

We zoom in on the primary BH's CSD in Figure \ref{fig:mdot_zoom} for the simulation with $a_b = R_H/4$.
Here, the density distribution is shown with filled contours, and the $\dot{M}$ shown is the radial (spherical) component of $\dot{M}$. 
In the $\dot{M}$ plot we also show a collection of the isodensity contours. 
The first thing to note is that the structure around the BH is not a classical "thin" accretion disk, rather it is a thick torus since the isodensity contours close on themselves \citep{Abramowicz.2004}. 
The second thing to note is that most of the accretion onto the BH proceeds either from directly above or through the surface layers of the torus. 
The interior of the torus is actually transporting mass away from the BH. 
This is facilitated by upwellings of material as gas that arrived from higher elevation is pushed outwards in the orbital plane. 
This motion then collides with another inwards flow of gas in the region between the BHs in the orbital plane. 
Material that is being transferred from the companion BH is thus deflected upwards and reaches the primary BH from the surface layers of the torus.

Figure \ref{fig:lprof} presents the specific angular momentum distribution of the torus and surrounding region. 
We plot both the total, angle-averaged, specific angular momentum profile in the top panel, and the $(R,z)$ distribution of the vertical component in the bottom panel. 
On the smallest scales near the accretion region, the torus has an approximately Keplerian rotation profile. 
Outside of $r \sim 0.15 a_b$, the angle-averaged specific angular momentum profile transitions to being sub-Keplerian both in magnitude and also in radial dependence. 
From Figure \ref{fig:mdot_zoom} and the bottom panel of Figure \ref{fig:lprof}, we see that the gas that eventually accretes onto the BH is made up of two populations in terms of angular momentum. 
There is very low angular momentum material that falls directly onto the BH, and there is higher angular momentum gas that accretes through the surface layers of the torus.  

At first glance, the fact that the CSDs surrounding each BH have outwards accretion near the mid-plane may be surprising. 
\rev{However, the AGN disk supplies gas to the CSD region from higher latitudes at a faster rate than the BHs can accrete (see for example the blue regions in the $\dot{M}$ panel of Figure \ref{fig:mdot_zoom}).
Thus in steady-state, there must be an outflow of gas near the CSD mid-plane to lower the overall $\dot{M}$ delivered to the BHs.   
It could be argued that this is a consequence of our chosen sink method and removal rate.
We leave the exploration of different sink parameters to a future work, however. 
}
\rev{Nevertheless}, this flow structure has also been seen in isothermal, 3D simulations of circum-planetary disks \citep[e.g.,][]{Tanigawa.2012}.
More recent simulations of these structures that include radiation, however, find that the mid-plane of the circum-planetary disk accretes inwards \citep[e.g.,][]{Judit.2014}.
Moreover, if the opacity near the planet is large, the gas heats up enough so that the circum-planetary disk transitions to a circum-planetary envelope \citep{Judit.2016}.
Future simulations of BH accretion on the scales of the Hill sphere that include a treatment for radiation may find similar behavior.

\subsection{Binary Evolution} \label{sec:adot}

In this Section we present steady-state torques and the rates at which the BBH orbital elements change with time. 
To study the orbital evolution of embedded BBHs, we use the procedures from the isolated binary literature \citep[see e.g.,][]{Munoz.2019,Zrake.2020knc}. 
The orbital angular momentum and orbital energy of the BBH are given by, 
\be
L_b &=& \frac{m_1 m_2}{m_b} \sqrt{G m_b a_b (1 - e_b^2)}  = \mu_b \ell_b ,\\ 
\mathcal{E}_b &=& - \frac{G m_b}{2 a_b} = \frac{1}{2} \dot{\vec{r}}_b^2 - \frac{G m_b}{r_b} ,
\ee
where in the first line we have defined the reduced mass $\mu_b = m_1 m_2 / m_b$ and the binary specific angular momentum, $\ell_b = \sqrt{G m_b a_b (1-e_b^2)}$. 
For a circular binary, $\ell_b = \sqrt{G m_b a_b} = n_b a_b^2$. 
The changes in $a_b$ and $e_b$ are related to the changes in $L_b$, $\mathcal{E}_b$, and $m_{1,2}$ via the expressions,
\be
\frac{\dot{a}_b}{a_b} &=& -\frac{\dot{\mathcal{E}}_b}{\mathcal{E}_b} + \frac{\dot{m}_b}{m_b} ,  \\ 
\frac{e_b\dot{e}_b}{1 - e_b^2} &=& - \frac{\dot{L}_b}{L_b} -\frac{\dot{\mathcal{E}}_b}{2 \mathcal{E}_b} + \frac{\dot{m}_1}{m_1} + \frac{\dot{m}_2}{m_2} .
\ee

Interaction with the disk adds additional angular momentum and energy to the BBH at the rates, 
\be
\dot{L}_{\rm ext} &=& r_1 \times \vec{f}_{\rm ext,1} + r_2 \times \vec{f}_{\rm ext, 2} \\ 
&=& \mu_b \vec{r}_b \times (\vec{a}_{\rm ext,2} - \vec{a}_{\rm ext,1} ) , \label{eq:torque} \\ 
\dot{\mathcal{E}}_{\rm ext} &=& \vec{v}_b \cdot (\vec{a}_{\rm ext,2} - \vec{a}_{\rm ext,1})  \nonumber \\
&-&  \frac{G \dot{m}_b}{r_b} + \frac{\dot{m}_1}{m_1} \vec{\dot{r}}_b \cdot \vec{\dot{r}}_1 - \frac{\dot{m}_2}{m_2} \vec{\dot{r}}_b \cdot \vec{\dot{r}}_2 \label{eq:power} ,
\ee
where $\vec{f}_{\rm ext,i} = d(m_i \dot{\vec{r}}_i)/dt$ and $\vec{a}_{\rm ext,i} = \vec{f}_{\rm ext,i}/m_i$\footnote{Note that our expression for the power differs from what is found in e.g. \citet{Munoz.2019}. This is because our accelerations are not the {\it specific} accelerations \rev{as defined in that paper, \rev{$d\vec{\dot{r}}_i/dt$}}, and include additional $\dot{m}$ dependencies. \rev{In fact, our $\vec{f}_{\rm ext}$ is what that paper defines as $d\Delta \vec{p}/dt$, where $\Delta \vec{p}$ is the change in momentum.}}.

\begin{figure}
\centering
\includegraphics[width=.48\textwidth]{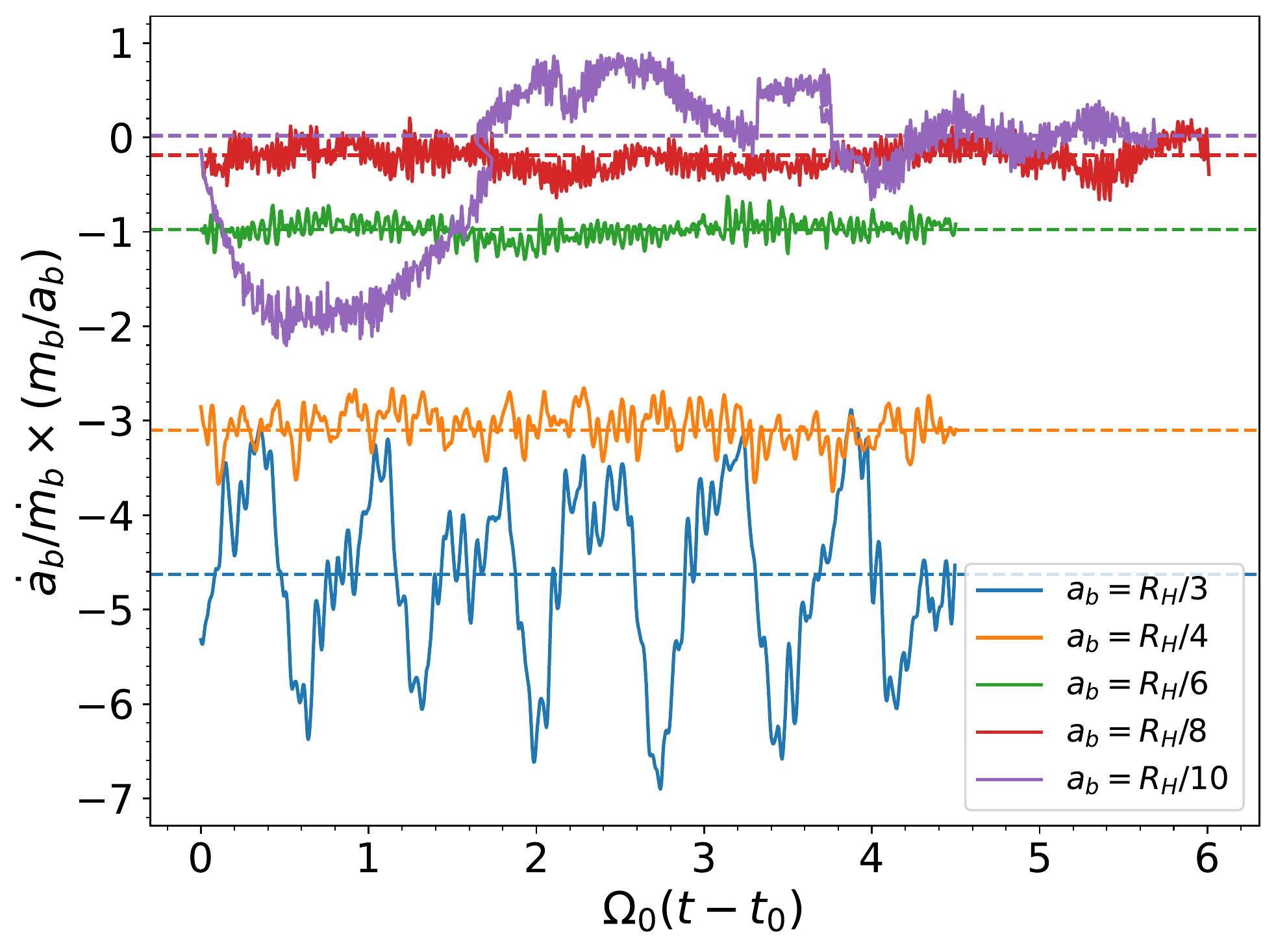}
\caption{Time evolution of the accretion eigenvalue $\dot{a}_b/\dot{m}_b$ \rev{at the end of each simulation. A window time-average of $\Omega_0 \Delta t =0.5$ has been applied to each point to smooth out the short time scale fluctuations. The final steady-state values for each simulation are shown with the dashed lines. All simulations reach a steady value. The values of $t_0$ for each simulation are $t_0=[15.5, 15.5, 15.5, 6.48, 0.5] \Omega_0^{-1}$ in order of decreasing $a_b$.} }
\label{fig:adot_time}
\end{figure}

\begin{table}[t]
\centering
\begin{tabular}{r| c c c c c}
$a_b/R_H$      & $1/3$	 & 	$1/4$	 & 	$1/6$	 & 	$1/8$	 & 	$1/10$ \\ \hline
$\ell_0^{g}$   & -1.85	 & 	-1.05	 & 	0.02	 & 	0.41	 & 	0.51 \\ 
$p_0^{g}$      & 3.60	 & 	2.08	 & 	-0.03	 & 	-0.81	 & 	-1.02 \\ \hline \hline 
$\ell_0^{acc}$ & 0.99	 & 	0.99	 & 	0.99	 & 	1.00	 & 	1.00 \\ 
$p_0^{acc}$    & 2.03	 & 	2.02	 & 	2.01	 & 	2.00	 & 	2.00 \\ \hline \hline 
$\ell_0$       & -0.86	 & 	-0.06	 & 	1.01	 & 	1.41	 & 	1.51 \\ 
$p_0$          & 5.63	 & 	4.10	 & 	1.98	 & 	1.19	 & 	0.98 \\ \hline \hline 
$\avg{\dot{a}_b / \dot{m}_b (m_b/a_b)}$   & -4.63	 & 	-3.10	 & 	-0.98	 & 	-0.19	 & 	+0.02 \\ 
$\avg{\dot{e_b^2} (m_b/\dot{m}_b)}$  & 0.10	 & 	0.01	 & 	-0.00	 & 	-0.01	 & 	-0.01
\end{tabular}
\caption{Final steady-state time-averaged torque, power, $\dot{a}_b$ and $\dot{e_b^2}$ values for each simulation. Each number is normalized to $\dot{m}_b$. The torque and power are split into gravitational and accretion components. 
\rev{Simulations with $a_b/R_H> 1/8$ are averaged over the whole time period shown in Figure \ref{fig:adot_time}, while the simulations with $a_b/r_H=1/8$ and $1/10$ are averaged over $\Omega_0 \Delta t=2.5$ and $=0.5$, respectively.}}
\label{tab:res}
\end{table}

\begin{figure}
\centering
\includegraphics[width=.48\textwidth]{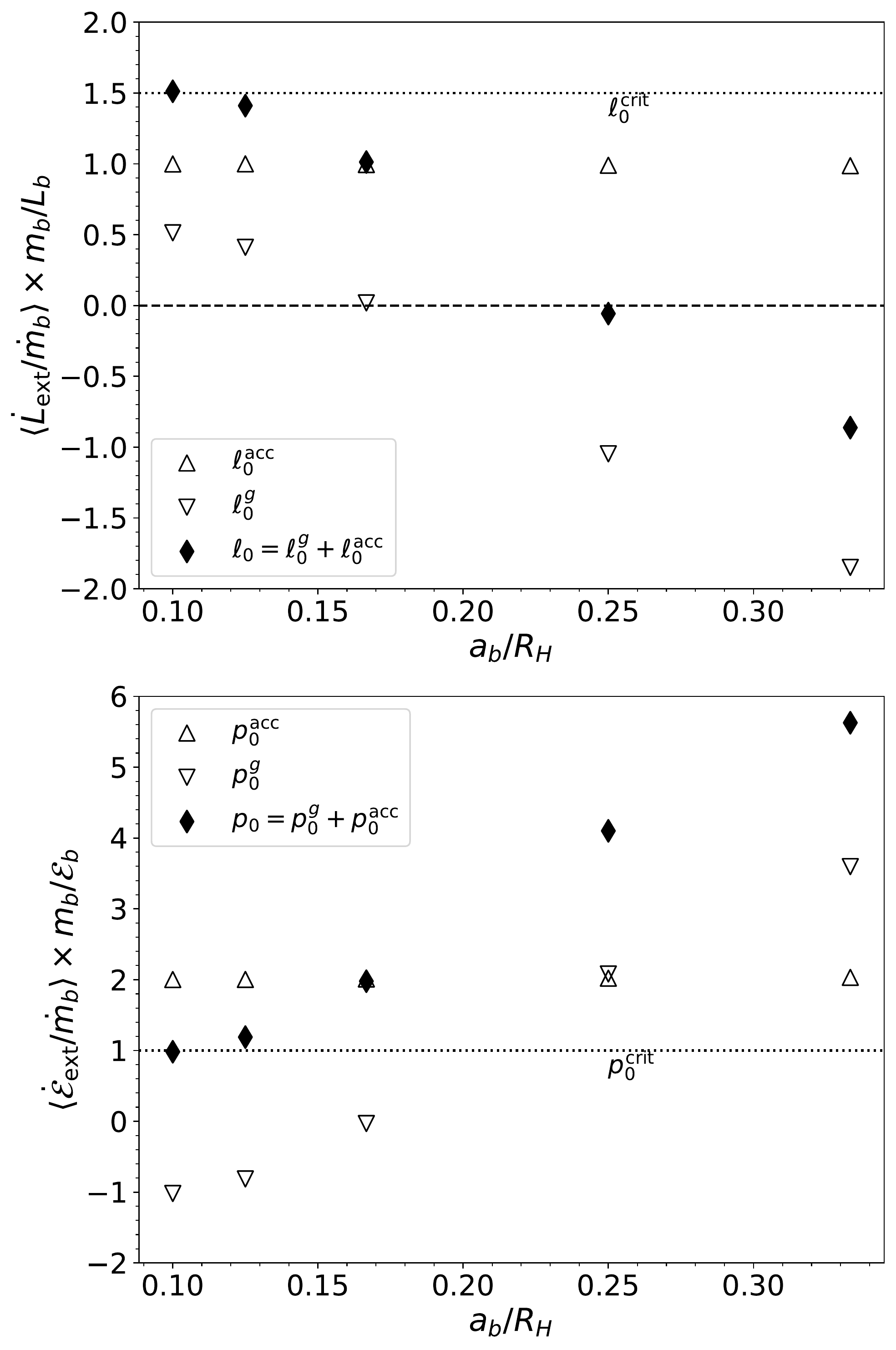} 
\caption{Final time-averaged torque values (top) and power values (bottom) as a function of binary separation. For each simulation we split the torque and power into gravitational (downward triangle), accretion (upward triangle), and total (filled diamonds) contributions.
All values are normalized to $\dot{m}_b/m_b$. 
The critical values where the binary expands are shown as dashed lines. 
The gravitational torque becomes positive for $a_b < R_H/5$, and exceeds the critical torque required for binary expansion at $a_b \sim R_H/10$.
}
\label{fig:trq_pwr}
\end{figure}
\begin{figure}
\centering
\includegraphics[width=.48\textwidth]{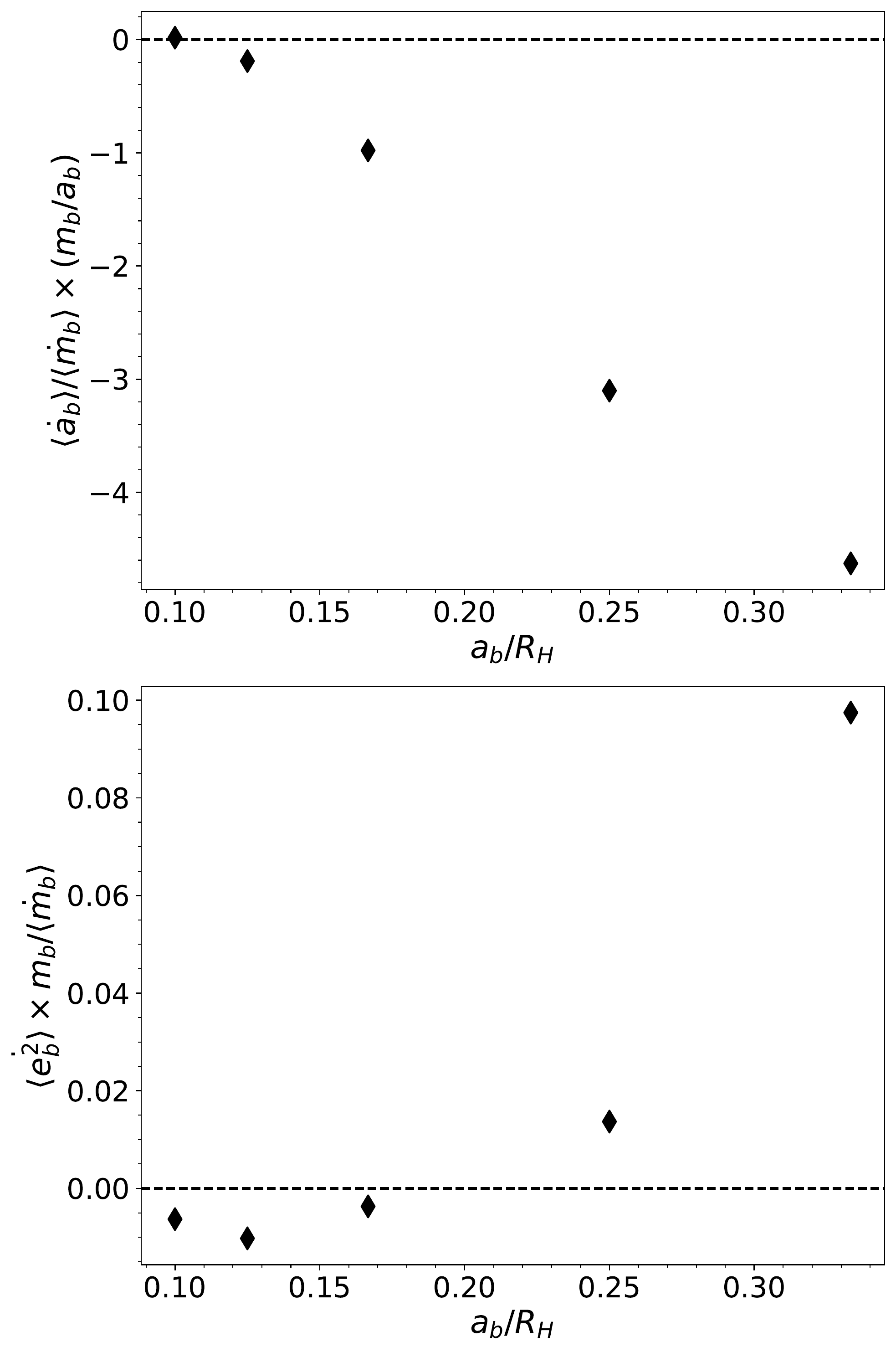} 
\caption{Final time-averaged $\dot{a}_b/a_b$ and $\dot{e_b^2}$ values as a function of binary separation. All values are normalized to $\dot{m}_b/m_b$. 
Binaries with $a_b > R_H/10$ contract, but with decreasing rates as the binary tightens. 
The critical binary separation below which we would expect an expanding binary is $\approx R_H/10$. 
Our widest binaries contract while growing their eccentricity. 
This is likely due to these binaries having an effective tidal eccentricity of $\gtrsim (a_b/R_H)^3$. 
}
\label{fig:adot}
\end{figure}

The external forces consist of both gravitational kicks and additional momentum gained through accreting gas.
We keep track of each force's contribution to the torque and power with the following definition. 
In Equations \eqref{eq:torque} and \eqref{eq:power} each $\vec{a}_{\rm ext}$ is split into a gravitational piece from the gravitational force and an accretion piece from the change in momentum induced by accretion. 
In addition, the terms proportional to $\dot{m}_b$ are added to the accretion contribution to $\dot{\mathcal{E}}_{\rm ext}$.

During each simulation we keep running time-averages of $\dot{L}_{\rm ext}$, $\dot{\mathcal{E}}_{\rm ext}$, and $\dot{m}_{1,2}$, and separately track the contributions from gravity and accretion. 
Using these we define the dimensionless quantities,
\be
\ell_0^g &=& = \avg{\frac{\dot{L}_{\rm ext,g}}{L_b} \frac{m_b}{\dot{m}_b}} , \qquad  \ell_0^{\rm acc} = \avg{\frac{\dot{L}_{\rm ext,acc}}{L_b} \frac{m_b}{\dot{m}_b}} , \\ 
p_0^g &=& \avg{\frac{\dot{\mathcal{E}}_{\rm ext,g}}{\mathcal{E}_b} \frac{m_b}{\dot{m}_b}} , \qquad  
p_0^{\rm acc} = \avg{\frac{\dot{\mathcal{E}}_{\rm ext,acc}}{\mathcal{E}_b}  \frac{m_b}{\dot{m}_b}} ,
\ee
and $\eta =  \avg{\dot{m}_2} / \avg{\dot{m}_1}$. 
Finally, we calculate the change in the orbital elements as,
\be
\avg{ \frac{\dot{a}_b}{a_b} \frac{m_b}{\dot{m}_b}} &=& 1 - p_0 , \label{eq:ad_avg} \\ 
\avg{ \dot{e_b^2} \frac{m_b}{\dot{m}_b}} &=& -2 \ell_0 - p_0 + 2  \left(\frac{(1+q_b)(1+\eta/q_b)}{1+\eta} \right) , \label{eq:ed_avg}
\ee
where e.g., $p_0 = p_0^g + p_0^{\rm acc}$ and $q_b=m_2/m_1$. 
Note that the last term in Equation \eqref{eq:ed_avg} is equal to $4$ for our $q_b=1$ binaries.
For circular binaries, $d(e_b^2)/dt \approx 0$, and we can substitute $p_0 = -2 \ell_0 + 4$ in Equation \eqref{eq:ad_avg}, 
\be
\avg{ \frac{\dot{a}_b}{a_b} \frac{m_b}{\dot{m}_b}} &=& 2 \ell_0 - 3  .
\ee
So long as $\ell_0 < 3/2$, the binary will contract. 

We find that $\dot{a}_b$ converges shortly after the Hill region has converged in time, i.e. after $\dot{M}_{\rm Hill} \approx0$. 
\rev{Figure \ref{fig:adot_time} shows the time-evolution of $\dot{a}_b$  over the last phase of the evolution. 
The data is smoothed over a time-averaging window of $\Omega \Delta t =0.5$ to remove the large amplitude oscillations  commonly seen in 2D simulations of isolated \citep[e.g.,][]{Munoz.2019} and embedded binaries \citep{Li.2022}.
Each simulation converges to a steady-state value (shown with the horizontal dashed lines). 
While simulations with $a_b \ge R_H/8$ show only the final AGN orbit, the results from the tightest binary actually show the entire simulation time\footnote{\rev{The number of time steps for a simulation to complete one AGN orbit scales roughly as the orbital period of gas at the sink radius. Since both the sink radius and the cell size scale with $a_b$, the number of steps scales as $\sim a_b^{-3/2}$. Thus, the $a_b = R_H/10$ simulation is roughly $\sim 6$ times more expensive (in wall clock time) than the $a_b=R_H/3$ simulation.}}.
For this case, the simulation has run just long enough to reach a steady-state value with a slightly positive $\dot{a}_b$. 
There is a clear trend of increasing accretion eigenvalue as the BBH separation decreases. 
}

Table \ref{tab:res} lists the steady-state values for $\ell_0$, $p_0$, $\dot{a}_b$ and $\dot{e_b^2}$. 
We plot the values of $\ell_0$ and $p_0$ versus $a_b/R_H$ in Figure \ref{fig:trq_pwr}, and split each point into their gravitational and accretion contributions. 
The dotted lines mark the critical values delineating expansion from contraction for these quantities. 
We find that there is a general trend of decreasing $p_0$ and increasing $\ell_0$ as the binary tightens. 
In fact, there is a torque reversal seen just below $a_b/R_H = 0.25$. 
Binaries with $a_b<R_H/4$ have positive torques, i.e. they gain angular momentum from the AGN disk, while wider binaries lose angular momentum to the disk. 

From Figure \ref{fig:trq_pwr}, the contributions to $\ell_0$ and $p_0$ from accretion are roughly independent of $a_b$, and are $\ell_0^{\rm acc} \approx 1$ and $p_0^{\rm acc} \approx 2$. 
This is expected when the accretion-induced {\it specific} torque and power are small, as we now show. 
Simplifying Equations \eqref{eq:torque} and \eqref{eq:power} for $m_1=m_2$ and a perfectly circular BBH, results in the following expressions for $\ell_0^{\rm acc}$ and $p_0^{\rm acc}$,
\be
\ell_0^{\rm acc} &=& \frac{m_b}{\dot{m}_b} \frac{ \vec{r}_b \times (\ddot{\vec{r}}_2 - \ddot{\vec{r}}_1)^{\rm acc}}{\ell_b} + 1 , \\ 
p_0^{\rm acc} &=& \frac{m_b}{\dot{m}_b} \frac{\vec{r}_b \cdot (\ddot{\vec{r}}_2 - \ddot{\vec{r}}_1)^{\rm acc}}{\mathcal{E}_b} + 2 .
\ee
The first terms in each equation depend on the specific binary acceleration, $\ddot{\vec{r}}_b = \ddot{\vec{r}}_2 - \ddot{\vec{r}}_1$, from accretion. 
We find that the contribution of these terms to the final rates is small, indicating that the details of the sink prescription are unimportant for our main conclusions.

In contrast to the accretion forces, the gravitational contributions to $\ell_0$ and $p_0$ show a clear linear trend with $a_b$. 
Wide binaries feel strong, negative torques and gain energy from the disk, whereas tight binaries gain angular momentum and lose energy to the disk. 

It should be noted that, even if a BBH gains angular momentum from the disk, it can still contract if $\ell_0 < 3/2$ and $p_0 > 1$. 
This is shown explicitly in Figure \ref{fig:adot}, where we plot $\avg{\dot{a}_b}$ as a function of $a_b$. 
Binaries wider than $\sim 0.1 R_H$ contract, while our tightest binary with $a_b = R_H/10$ expands its orbit.

The bottom panel of Figure \ref{fig:adot} shows the measured values of $\avg{\dot{e_b^2}}$. 
The expectation is that circular binaries should remain circular with small, negative values of $\dot{e_b^2}$. 
Our simulations agree on this for binaries with $a_b < R_H/5$, but find that wider binaries grow their eccentricity. 
As mentioned in Section \ref{sec:nbody}, tidal interaction with the SMBH induces an eccentricity of $e_b \sim (a_b/R_H)^3$, which for $a_b=R_H/3$ is $\sim 0.04$.
The osculating eccentricity for this simulation oscillates around a time-average of $e_b \approx 0.05$, with a peak value of $e_b \sim 0.1$ (as shown by the orange line in Figure \ref{fig:nbody}). 
Simulations of isolated binaries have found that binaries with $e_b\gtrsim0.07$ increase their eccentricity \citep{Munoz.2019,DOrazio.2021}, and so our widest binary may be increasing its eccentricity for similar reasons.

\subsection{Torque profiles} \label{sec:torques}

\begin{figure*}[t]
\centering
\includegraphics[width=.48\textwidth]{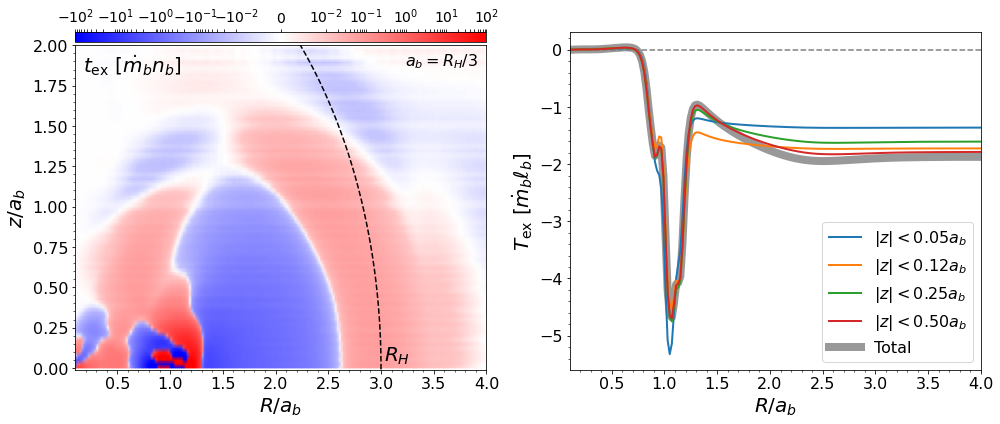} 
\includegraphics[width=.48\textwidth]{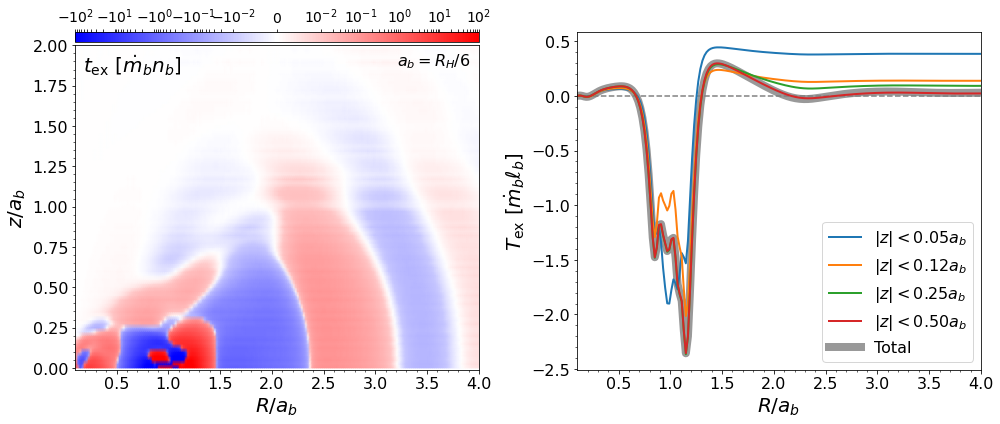}  \\
\includegraphics[width=.48\textwidth]{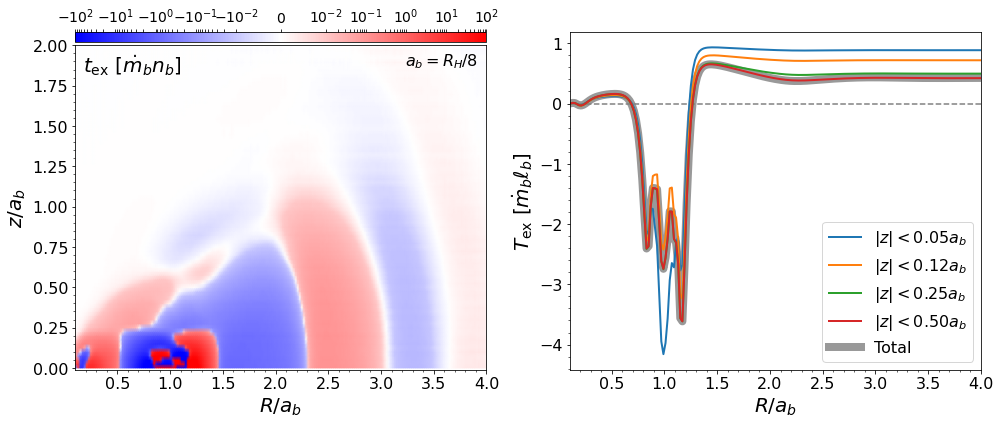} 
\includegraphics[width=.48\textwidth]{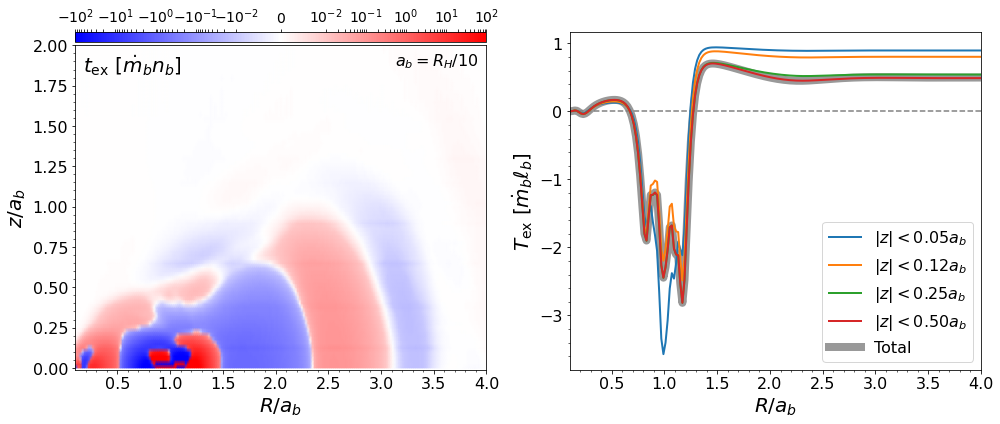} 
\caption{Final time- and azimuthally-averaged $t_{\rm ex}(R,z)$ profiles (left plots) and cumulative torque $T_{\rm ex}=\int_0^r \int_{-z_i}^{z_i} t_{\rm ex} dz dR$ integrated over different heights (right plots) for four simulations. 
The $t_{\rm ex}$ plots show the symmetric torque density, i.e., $(t_{\rm ex}(R,-|z|) + t_{\rm ex}(R,+|z|))$.  
The mid-plane torque is very asymmetric except for the tightest binary which has a substantial amount of positive torque from $1.3<r/a_b<1.5$. Most of the asymmetry across the perturbing BH's orbit comes off mid-plane at heights $|z|<0.25$. These heights are also where most of the negative outer torque comes from at $r/a_b>1.3$. 
\rev{The horizontal white lines in the $t_{\rm ex}(R,z)$ plots are artifacts of the averaging and plotting process and do not affect any of our conclusions.}
}
\label{fig:tex}
\end{figure*}

A useful way to understand the origins of the $\dot{a}_b$ values shown in Figure \ref{fig:adot} is to examine the spatial distribution of the torque of the gas on the binary.  
Torque maps are commonly employed in analysis of planet-disk and binary-disk interaction \citep{Dempsey.20204f,Munoz.2019,Munoz.2020,Li.2021,Li.2022}. 
For circular binaries, the torque directly determines $\dot{a}_b$, but for eccentric binaries, it is the power that determines $\dot{a}_b$. 
Nevertheless, the torque still approximately determines $\dot{a}_b$, and so in this Section we focus on it instead of the spatial distribution of the power.

We generate time and azimuthally averaged torque maps with the procedure outlined in Section \ref{sec:profiles} .
The torques are computed during each kick step in the simulation by keeping running time averages of the two torque densities \citep{Li.2022},
\be
t_{\rm ex,1} &=& (\vec{r}-\vec{r}_2) \times \frac{\vec{r}-\vec{r}_1}{|\vec{r}-\vec{r}_1|^3} G \rho ,\\ 
t_{\rm ex,2} &=& (\vec{r}-\vec{r}_1) \times \frac{\vec{r}-\vec{r}_2}{|\vec{r}-\vec{r}_2|^3}  G \rho .
\ee
The sum $\int (t_{\rm ex,1} + t_{\rm ex,2}) dV $ is equal to the total torque acting on the binary separation. 
Note that, e.g. $t_{\rm ex,2}$ can be thought of as the torque of the disk on BH 2 computed in a coordinate system centered on BH 1.
When expressed this way, we can draw a direct parallel to planet-disk interactions where one BH is the "star" (i.e the center of the coordinate system) and the companion BH is the "planet" (i.e the perturber). 
When discussing the torques, we will refer to the BH at the center of the coordinate system as the primary, and the companion BH as the secondary. 
Because we focus on circular equal-mass binaries, $t_{\rm ex,1} \approx t_{\rm ex,2}$ as the problem is symmetric \citep{Li.2022}.

Figure \ref{fig:tex} shows the $t_{\rm ex}(R,z) = t_{\rm ex,1}(R,z) + t_{\rm ex,2}(R,z)$ profiles for each simulation. 
Because the results are very close to vertically symmetric, we add the bottom half (i.e $z<0$) of the domain to the top half and plot the result in the top half. 
Red regions correspond to regions of the disk that torque the BH up, while blue corresponds to regions where the disk torques the BH down.
The perturbing BH, i.e. the one we are measuring the torque on, is at $R=a_b$ and $z=0$. 
Along with each 2D distribution of $t_{\rm ex}$, we add an additional plot showing the height-integrated $t_{\rm ex}$ profiles, which in addition to being integrated in $z$, are also integrated in $R$ in a cumulative sense, i.e we plot $T_{\rm ex} = \int_0^R \int_{-z_i}^{z_i} t_{\rm ex}(R',z) dz dR'$ for a chosen value of $z_i$. 
Regions where these curves are flat do not torque the binary, and the values far from the binary measure the total torque on the BH in the specified z-ranges.

In \citet{Li.2022}, we identified three major regions that torque the BBH using 2D simulations: the primary's CSD ($R<0.5 a_b$), the secondary's CSD  ($0.5 a_b<R<1.5 a_b$), and the CBD ($R>1.5 a_b$). 
In that work, the torque excited in the secondary's CSD typically determined the overall sign of the torque on the BBH. 
In 3D, each region has a $z$ dependence in addition to its $R$ dependence. 
Using the $t_{\rm ex}(R,z)$ and $\rho(R,Z)$ profiles as a guide, we use the following definitions for each region: the primary BH CSD is the region where $R<0.5 a_b$ and $|z| < 0.25 a_b$; the companion BH CSD is the region where $0.5 a_b < R < 1.5 a_b$ and $|z|<0.25 a_b$; and the CBD is made up of two regions, one at $R>1.5 a_b$ and $|z|>0$ and one at $R>0$ and $|z|>0.25 a_b$. 
We now discuss the torque from each region in detail.

In the primary BH's CSD, the companion BH raises spiral waves that tend to torque it (and the binary) up. 
These spirals are similar to the inner-disk spirals seen in planet-disk interaction and predominantly add angular momentum to the BH. 
From the $T_{\rm ex}$ curves, the torque within $R < 0.5 a_b$ comes mostly from the orbital plane and becomes stronger as $a_b$ decreases. 

Similar to the primary BH's CSD, the secondary BH's CSD also contains spiral waves that torque the companion. 
These waves are excited by tidal interaction with the primary BH.
The material in this region exhibits the strongest variation with binary separation -- likely because it is the closest to the companion BH and thus the most strongly interacting. 
Whether this region torques the binary up or down depends on the asymmetry across the CSD, as the inner half tends to torque the binary down, while the exterior half wants to torque it up. 
In all but the widest binary, the exterior side has a stronger torque and the net result is that the companion CSD adds a net positive amount of angular momentum to the binary. 
And much like the primary BH's CSD, most of the torque from the companion BH's CSD is confined to near the mid-plane.

Finally, we turn to the CBD.  
The part of the CBD directly above the binary at $|z|>0.25 a_b$ imparts a relatively small torque, as there is not much difference between the red and green curves in Figure \ref{fig:tex} at $R<1.5 a_b$. 
The only noticeable difference between the two curves occurs outside of $R\sim 1.5 a_b$.  
Indeed, this part of the CBD is vertically extended and follows the width of the CBD spirals seen in the right panels of Figure \ref{fig:gas_bbh}. 
The near side of this region corresponds to the nearest outer spiral to the companion and torques the binary down, while the far side corresponds to the farther outer spiral and torques the binary up \citep[][]{Li.2021,Li.2022}.
From the $T_{\rm ex}$ curves we see that the near spiral torques the BH far more than the outer spiral, resulting in a net negative CBD torque. 
And, moreover, most of the torque comes from off of the orbital plane due to the vertically extended CBD spirals. 

Overall, when decreasing the binary separation, two things occur. 
One is that the torque from both BHs' CSDs increases to the point of becoming positive. 
And second is that the torque from the CBD becomes weaker.
In fact, the CBD weakens to the point that at just below $a_b \sim R_H/4$, the total torque on the binary switches sign, and at $a_b \sim R_H/10$ the total torque becomes large enough to result in $\avg{\dot{a}_b} > 0$. 

The result that tighter embedded BBHs should have positive torques is not too surprising given the recent findings of the isolated binary community \isolatedp{e.g.,}. 
In some sense, when $a_b \ll R_H$, the fact that the binary is embedded in a background accretion disk becomes less important. 
On these scales, the gas essentially only feels the gravitational effects of the binary and so we would expect our results to become more consistent with an isolated binary (i.e., expansion).
This analogy is not perfect, however, as the background AGN disk still delivers mass to the binary region in a completely different manner to how mass is delivered to an isolated binary. 
In the isolated case, there is a large viscous accretion disk that slowly builds the CSDs in a mostly 2D manner (if the binary is aligned with the CBD). 
But, as shown in the previous Section, the CSD around a BH in an AGN disk is fed in a predominately 3D manner and develops into a torus. 
Moreover, the CBD is not necessarily viscously controlled or even large in extent as it is bounded between $r\sim a_b$ and $r\sim R_H$. 
Nevertheless, despite the different accretion processes, when the torques are normalized to $\dot{m}_b \ell_b$, the numbers are qualitatively similar to the isolated binary results. 

\section{Discussion} \label{sec:discussion}

In this Section, we explore a few variations on our standard setup, put our accretion values in the context of the Eddington rate, and discuss some of the future directions for simulations of embedded BBHs.

\subsection{Accretion rates} \label{sec:rates}

We compare our values of $\dot{m}_b$ to two values: the Bondi rate \citep{Bondi.1952,Edgar.2004}, and the Eddington rate. 
The Bondi rate  is defined as, 
\be
\dot{M}_{\rm Bondi} \approx 4 \pi R_b^2 \rho c_s ,
\ee
where $R_b = G m_{b}/c_s^2$ is the Bondi radius. 
In terms of the Hill radius, $\dot{M}_{\rm Bondi}$ can be expressed as, 
\be
\dot{M}_{\rm Bondi} \approx \frac{4 \pi}{3} R_{\rm H}^3 \rho \left(\frac{3 R_H}{H} \right)^3 \Omega_0 .
\ee
Approximating $M_d \sim 4 \pi R_H^3 \rho /3$, results in the Bondi accretion rate of $\dot{M}_{\rm Bondi}/m_b \sim 13.8 \Omega_0$ for $M_d=m_b$ and $R_H/H=0.8$. 
This number is much larger than the $\avg{\dot{m}_b}$ values obtained in our simulations which are between \rev{$\approx [0.15,0.34] m_b \Omega_0$}. 
Evidently, the CSD can inhibit the accretion flow from the purely spherical Bondi rate by a factor of $\gtrsim 50$. 

The Eddington rate is a nearly universal value of $\dot{M}_{\rm edd}/m_b  \approx 7 \times 10^{-16} \, {\rm s}^{-1}$. 
To get some sense of how this number compares to our results, suppose that our BBH has a total mass of $60 M_\odot$ and is separated by $10^7 r_g$, where $r_g = G m_b/c^2$ is the gravitational radius of one of the BHs. 
This BBH has a mean motion of $n_b \sim 10^{-7} \, {\rm s}^{-1}$ and, if we take our $a_b=R_H/6$ result for $\dot{m}_b$, a measured $\dot{m}_b/m_b \sim 10^{-6} {\rm s}^{-1}$. 
This accretion rate, as well as the Bondi rate, is $\sim 10$ orders of magnitude faster than the Eddington rate. 
Our estimate was scaled to a Hill sphere mass equal to the binary mass, though, so lighter disks can bring this number down.
Using Figure \ref{fig:agn} as a rough guide, $M_d/m_b$ may reach values $<10^{-4}$. 
This would bring $\dot{m}_b$ to within $\sim 6$ orders of magnitude of the Eddington rate. 
Moreover, moving the BBH outwards in the AGN disk will also lower $n_b$, but it is clear that the disk accretion onto the BH --at least on the scales we resolve -- is hyper-Eddington.

\subsection{Embedded-ness} \label{sec:diffh}
\begin{figure}
    \centering
    \includegraphics[width=.48\textwidth]{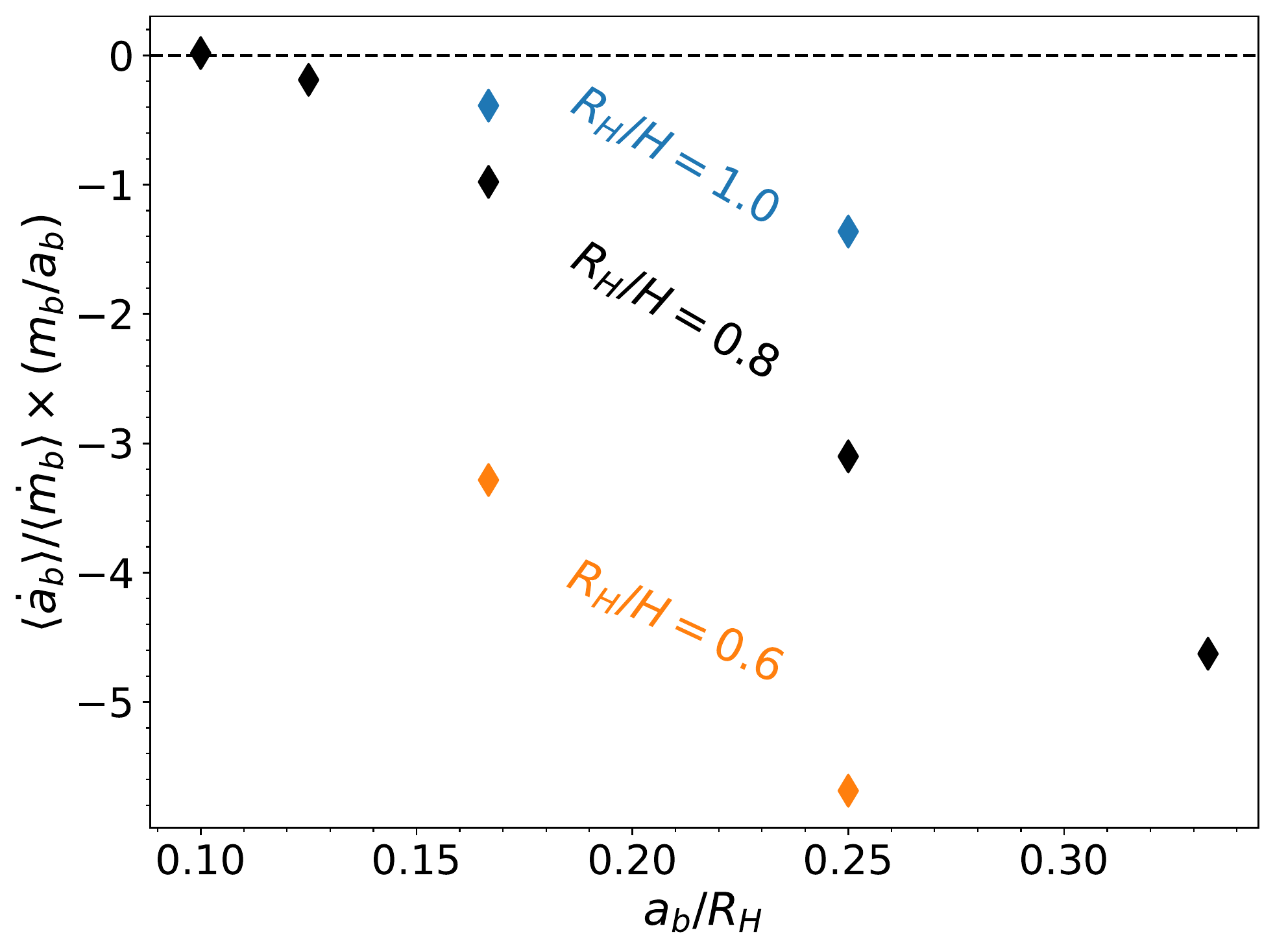}
    \caption{Accretion eigenvalues for different $R_H/H$ values. Orange, black, and blue points are for $R_H/H=0.6, 0.8,$ and $1.0$, respectively. As BBHs become more embedded (lower $R_H/H$), they contract at a faster rate. The critical separation where BBHs switch from contraction to expansion thus decreases as $R_H/H$ decreases.}
    \label{fig:diffh}
\end{figure}

For computational reasons, we have restricted ourselves to just one value of $R_H/H$. 
However, as shown in Figure \ref{fig:agn}, we expect there to be a large range of plausible values of this parameter throughout an AGN disk. 
Recently, \citet{Kaaz.2021crx} have demonstrated that at small values of $R_H/H$ the Bondi radius can become comparable to the size of the BBH orbit. 
In this situation, the ram pressure of the gas prevents the formation of CSDs and accretion proceeds in a Bondi-like fashion while the binary contracts. 
But, even if a BBH starts at a small value of $R_H/H$, the mass growth will increase $R_H$, and if the time scale for this growth is faster than the time scale for the binary to merge, the CSDs may have time to grow and stall the binary at $R_H \sim H$. 
Any feedback from the BBH may counteract this, though, by either increasing the local value of $H$ and/or disrupting the CSDs.

To get a sense of how the transition point in $\avg{\dot{a}_b}$ depends on $R_H/H$, we have run additional simulations at $R_H/H=0.6$ and $R_H/H=1.0$ for binaries with $a_b/R_H=1/6$ and $1/4$. 
Other than the value of $R_H/H$, the simulations are identical to our main set of simulations. 

Figure \ref{fig:diffh} compares the steady-state $\avg{\dot{a}_b}$ for these new simulations to the results shown in Figure \ref{fig:adot}. 
We find that the contraction rate becomes more negative as the binary becomes more embedded.
This means that the separation at which the binary expands increases as $R_H/H$ increases. 
Furthermore, this agrees with our expectation that more embedded binaries have weaker CSDs due to a smaller Bondi radius.
\rev{Recall that CSDs act to expand the BBH orbit through the torque excited by the spirals in the CSDs, whereas the spirals in the CBD act to contract the orbit.
Since weaker CSDs produce weaker spirals (and hence weaker positive torques on the binary), we would expect more negative values of $\dot{a}_b$ as $R_H/H$ decreases.}

\subsection{Comparison to 2D} \label{sec:2d}

\begin{figure}
    \centering
    \includegraphics[width=.48\textwidth]{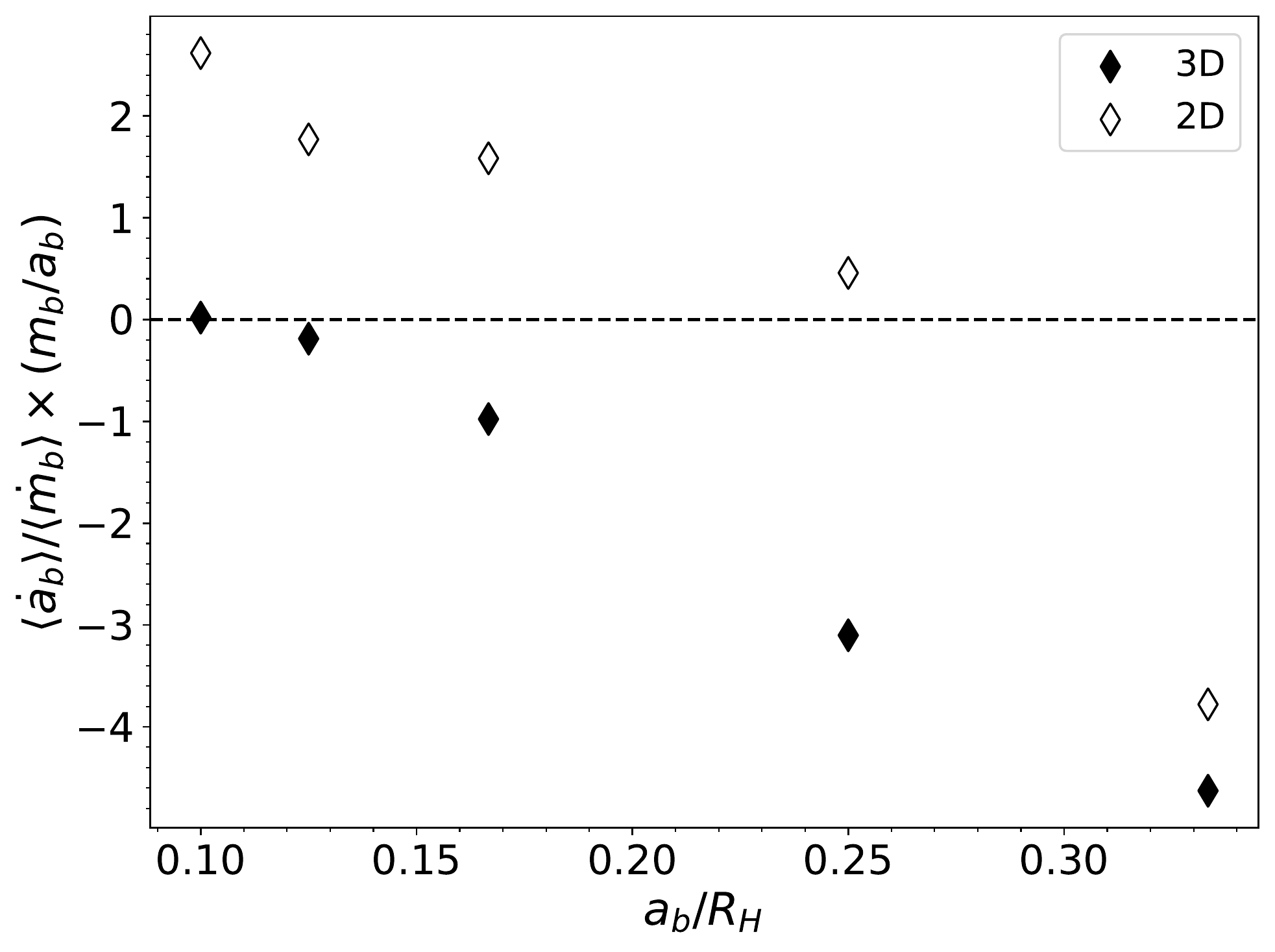}
    \caption{Comparison of the accretion eigenvalue between 3D (filled diamonds) and 2D (open diamonds) simulations. The 3D points are taken from Figure \ref{fig:adot}. In general, 2D simulations find larger values of $\avg{\dot{a}_b/\dot{m}_b}$. Both methods agree that binaries with $a_b=R_H/3$ contract, but disagree on where the transition from contraction to expansion is. }
    \label{fig:2d}
\end{figure}

\begin{figure}
    \centering
    \includegraphics[width=.48\textwidth]{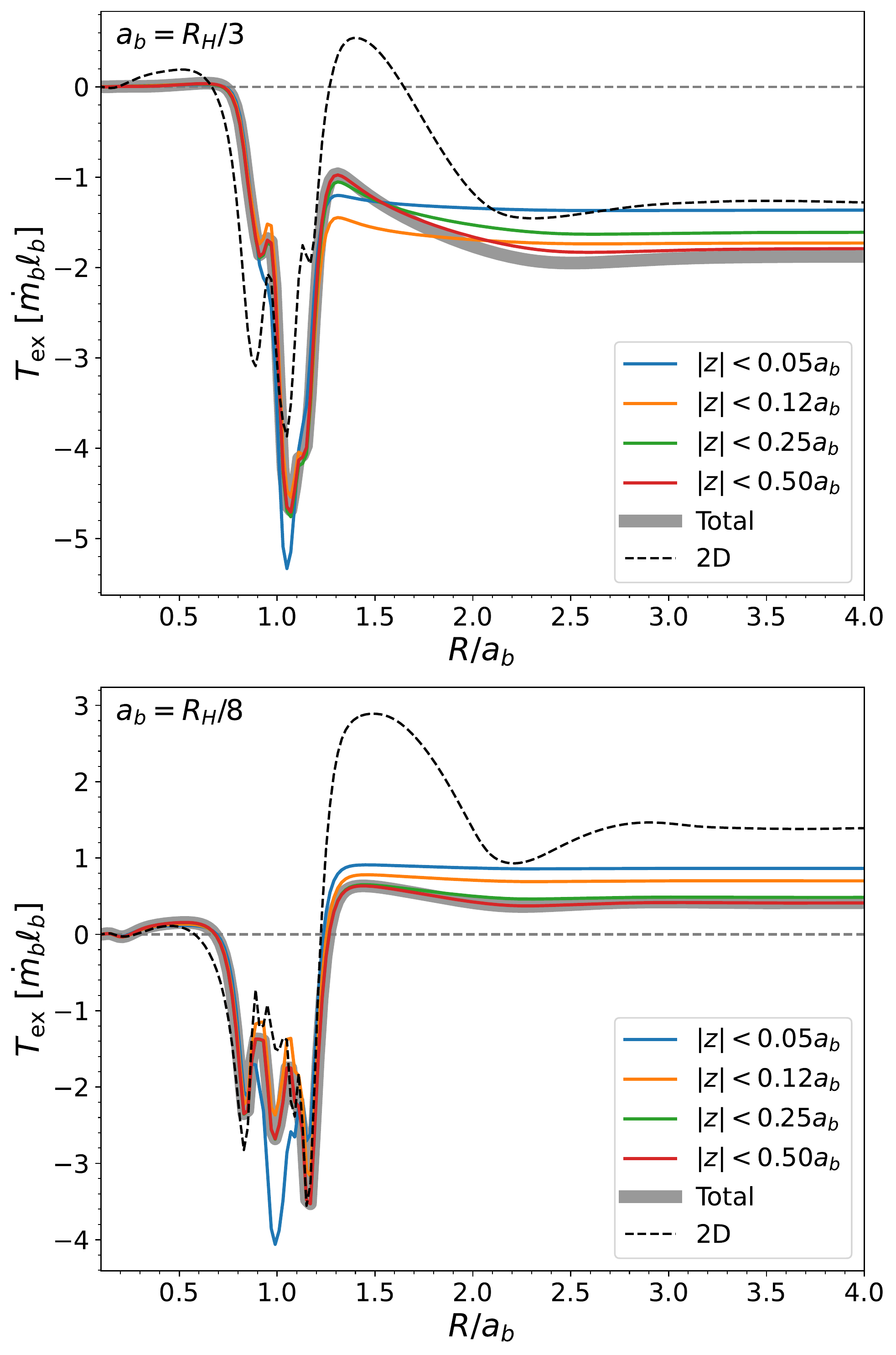}
    \caption{Comparison of the 2D and 3D $T_{\rm ex}$ profiles for two BBHs with $a_b=R_H/3$ and $R_H/8$. The 3D $T_{\rm ex}$ profiles are taken from Figure \ref{fig:tex}. The 2D simulations find much stronger CSD torques compared to the 3D simulations. }
    \label{fig:2d_torque}
\end{figure}

As mentioned in Section \ref{sec:intro}, almost all of the previous work on hydrodynamical simulations of embedded BBHs in AGN disks has been restricted to two dimensions. 
It is thus worth comparing our 3D results to equivalent 2D simulations. 
To do this, we run each of our simulations in the 2D shearing-sheet with the same resolution, refinement regions, and softening scale.

Figure \ref{fig:2d} compares the steady-state 2D \rev{$\dot{a}_b/\dot{m}_b$} values to the 3D results that were shown in Figure \ref{fig:adot}. 
In general, the 2D approximation results in too much positive torque, which leads to expanding binaries for all but our widest separation. 

To understand why this occurs, we compare the 2D and 3D radial torque profiles in Figure \ref{fig:2d_torque}. 
For clarity, we focus on just two separations $a_b/R_H=1/3$ and $1/8$. 
Comparing the value of the torque at $R=4 a_b$, we see that the 2D total torque is relatively close to the total mid-plane torque of the 3D simulation. 
This is despite the fact that the 2D CSD imparts much more positive torque onto the binary than the 3D CSD. 
The large 2D CSD torque is balanced by a much larger CBD torque that brings the total 2D torque value closer to the 3D mid-plane value. 
What the 2D simulations seem to be missing, however, is the extra negative CBD torque generated at higher altitudes that results in an overall lower 3D torque.

Recently, \citet{Lai.2022} ran 2D, adiabatic, shearing-sheet simulations of embedded BBHs including equal-mass, circular binaries. Compared to our setup, they focused on wider binaries with $a_b/R_H = 0.28$ and $0.58$ and $R_H/H = 0.7$. 
But, unlike our simulations, they did not include the tidal force of the SMBH on the BBH \rev{separation}\footnote{ \rev{ \citet{Lai.2022} did include the precession induced by the SMBH on their eccentric BBHs.}}.
At their large binary separations, the tidal force should induce a significant eccentricity, and possibly even disrupt the $a_b = 0.58 R_H$ binary completely \citep[cf. Figure \ref{fig:nbody} and][]{Eggleton.1995,Mardling.2001,Li.2021}. 
Their 2D accretion eigenvalue for $a_b/R_H= 0.28$ is $\sim 25\%$ more negative than our 2D isothermal $R_H/H=0.8$, $a_b/R_H= 1/3$ value, but is close to our 3D value. 
The fact that \citet{Lai.2022} find stronger negative torques is not too surprising given our results at smaller $R_H/H$ (cf. Figure \ref{fig:diffh}) and the results of \citet{Li.2022} which showed that hotter CSDs result in weaker torques that can lead to binary contraction. 
Since the temperature is allowed to vary in \citet{Lai.2022}, presumably the CSDs are hotter than our isothermal results, and thus the torques should be more negative.

\subsection{Extensions} \label{sec:extension}

In this section, we discuss several ways in which one could improve upon the simulations presented in this paper.

\paragraph{Orbital properties}

We have focused exclusively on circular, equal mass ratio BBHs. 
But, the binary formation process almost certainly results in eccentric, non-equal mass, and inclined BBHs -- especially if they are formed via capture from a circum-nuclear cluster of BHs \citep[e.g.,][]{Ostriker.1983,Syer.1991,Artymowicz.1993,McKernan.2011} as opposed to {\it in-situ} formation via gravitational instability \citep[e.g.,][]{Goodman.2004,Thompson.2005}.
Future work will explore the effects of eccentricity, mass ratio, and inclination with similar 3D simulations to the ones that we have presented here. 
Even without running these simulations, however, we already have strong expectations for the results from past studies on isolated binaries. 

In particular, we expect there to be a transition from expansion to contraction when a tight binary becomes eccentric \citep{Munoz.2019,DOrazio.2021} and when the mass ratio decreases \citep{Duffell.2020,Dempsey.2021}.  
Inclination is less explored in the isolated binary literature \citep[with the exception of][]{Moody.2019}, but \citet{Li.2021} found that in 2D, retrograde BBHs contract while prograde BBHs expand, and so we expect there to be a critical inclination above which an expanding binary would contract. 

Because the size of the orbit of BBHs is relatively small compared to the disk scale height, there is no {\it a-priori} constraint on how inclined these binaries can become. 
The binary formation mechanism should set some inclination distribution. 
Given this, embedded BBHs that are both eccentric and inclined could satisfy the criteria for Kozai-Lidov oscillations \citep[][]{Kozai.1962,Lidov.1962,Naoz.2016}. 
This could in principle drive the binary to very large eccentricities and help facilitate merger. 

\paragraph{Self gravity}

Depending on what the BBH-SMBH separation is, self-gravity in the AGN disk may be dynamically important. 
In fact, many AGN disk models find that the outer regions of the AGN disk ought to be close to Toomre unstable. 
Moreover, self-gravity may play an important role in the tori surrounding each BH. 
Self-gravity will also have an impact on the spirals that torque the binary, possibly weakening the strength of the CSD torques.

\paragraph{Magnetic fields}

One major component missing from our hydrodynamical simulations are magnetic fields. 
Unlike the mid-planes of protoplanetary disks, AGN disks are expected to be highly ionized and prone to the magneto-rotational instability (MRI, also known as Velikhov-Chandrashekhar instability; \citealt{Balbus91, Jiang20}).
This begs the question of how our simple hydrodynamic picture of the BBH region changes if it were to be embedded in an MRI active region of the AGN disk. 
We envision at least two things changing from our simple hydrodynamical picture. 
For one, strong turbulence driven by the MRI may disrupt the CSDs enough to severely weaken the torques from these regions and drive fast coalescence of the binary. 
In addition, any initially weak poloidal magnetic flux may be amplified on the scale of the CSDs and possibly drive strong jets \citep[e.g.,][]{McKinney06} or winds \citep[e.g.,][]{Giustini19} .

\paragraph{Radiation}

Radiation is often a critical component of AGN disk models \citep[e.g.,][]{Sirko.2003,Thompson.2005,Dittmann.2019}. 
It can provide additional vertical pressure to stabilize the disk and heat the gas to high temperatures. 
Radiation is bound to become even more important when examining the region around an accreting (and thus radiating) BH \citep[][]{Mishra16, Jiang20}. 
In fact, if the radiation in the vicinity of the BH is strong enough, detailed radiation transport from small scales up to the AGN disk scale may be necessary to obtain an accurate picture of the dynamics.  
Whether or not radiation is important depends on the accretion rate onto the BH, the disk density, and the opacity profile of the disk.
These properties will also set whether simplified treatments of radiation suffice, or whether full transport is required. 

\section{Conclusion} \label{sec:conclusion}

Using 3D shearing-box simulations we have examined the steady-state structure surrounding and torque acting on a BBH embedded in an AGN disk. 
Our main results are as follows:

\begin{itemize}

\item Depending on the AGN disk model and where the BBH is located in the disk, the ratio $R_H/H$ can easily range from $\ll 1$ to $\gg 1$ (Figure \ref{fig:agn}).
Moreover, the binary is typically more massive than the local gas if it is inside of $10^5 R_g$ of the SMBH.
Thus, our assumption of a fixed binary is not unjustified, as the timescale to change the binary separation may be longer than the timescale to enter a quasi-steady-state.

\item 3D accretion builds up a thick, sub-Keplerian torus around each BH that extends out to $\sim 0.5 a_b$ and a thick CBD that extends out to $\sim R_H$ (Figures \ref{fig:gas_bbh}, \ref{fig:mdot_2D_prof}, \ref{fig:mdot_zoom},  and \ref{fig:lprof}). 
In each of these disks, there are tidally excited $m=2$ spirals, in addition to a large scale high-$m$ spiral exited by the BBH center of mass at distances of $\sim H$.
On scales larger than $R_H$, the flow is similar to the flow around a single BH with total mass equal to the BBH mass (Figure \ref{fig:gas_single}). 

\item Each BBH accretes at a rate of between \rev{$\dot{m}_b \sim [0.15-0.34] m_b \Omega_0$} when we normalize the mass within the Hill sphere to the binary mass (Table \ref{tab:mdot}). 
This accretion rate, and the accretion rate of gas through the Hill sphere, stabilizes after a relatively short amount of time, $\Omega_0 t \sim$ few (Figures \ref{fig:mdot_time} and \ref{fig:mdot_profs}). 
Accretion onto each BH mostly occurs through the surface layers of the CSDs and through low angular momentum material (Figure \ref{fig:mdot_zoom}).
The $\dot{m}_b$ values we find are less than the Bondi rate, but many orders of magnitude faster than the Eddington rate. 

\item Our main result is the steady-state values of $\avg{\dot{a}_b}$ and $\avg{\dot{e_b^2}}$ shown in Figures \ref{fig:trq_pwr}~and~\ref{fig:adot} and Table \ref{tab:res}.
We find a consistent, separation-independent contribution to these rates from accretion, indicating that our sink prescription induces a very small torque to the binary. 
Additionally, as the binary separation decreases, the gravitational torque on the binary increases from a negative value to a positive value in a roughly linear manner with $a_b/R_H$.
This is driven by the strengthening of the positive CSD torques that eventually overwhelm the progressively weaker (and negative) CBD torques (Figure \ref{fig:tex}). 
In all but our widest binaries, we find that circular BBHs remain circular. 

\item We find that 3D simulations tend to have more negative torques than their 2D counterparts. This primarily arises from a weaker positive torque in the CSD region, as well as a more negative torque from a vertically extended CBD region (Figure \ref{fig:2d_torque}). 
This results in the critical $a_b$ below which the binary expands decreasing from $a_b \lesssim 0.3 R_H$ in 2D to $a_b \lesssim 0.1 R_H$ in 3D (Figure \ref{fig:2d}). 
Additionally, this critical point increases to wider separations as $R_H/H$ increases (Figure \ref{fig:diffh}). 

\end{itemize}

The results shown in Figure \ref{fig:adot} suggest that there is an attraction point in ($a_b/R_H,R_H/H$) space where any BBH will eventually stall at $a_b\approx 0.1 R_H$ while continually growing in mass.
We stress, however, that our main results are only valid for one value of $R_H/H$. 
As the BBH grows in mass, the value of $R_H/H$ increases. 
Therefore, to determine if there is, in fact, an attractor, one should explore other values of $R_H/H$ or include the evolution of the binary in the simulation. 
Our preliminary simulations at other $R_H/H$ values suggest that this critical point moves to larger $a_b$ as $R_H$ increases. 
This suggests that an initially wide binary that contracts while increasing its $R_H$ through mass accretion will eventually stall as it encounters the critical separation value.
Once in this attraction point, the binary will continue to increase its mass at fixed separation.

One way around this stalled evolution, however, is to embed the binary in a more massive AGN disk. 
If the surrounding gas has a large mass relative to the BBH, the $\dot{a}_b$ induced by the disk may be so fast that there is not enough time for the CSD torques to reach their steady-state value before the binary passes the critical separation.
In this scenario, an initially wide binary may contract all the way down to the horizon scales, all whilst growing in mass. 
Moreover, once $R_H>H$ the BBH may open a gap in the AGN disk. 
This could temporarily\footnote{On long timescales deep gaps do not inhibit accretion as shown by many recent simulations of high mass planet-disk interaction \citep[e.g.,][]{Dempsey.20204f,Dempsey.2020,Dempsey.2021}} lower the $\dot{M}$ feeding the BHs and possibly alter the torque profiles. 
But from the AGN models presented in Figure \ref{fig:agn}, only BBHs in the far outer regions of the AGN disk may be small in mass compared to their surroundings.

Perhaps a more likely scenario, given the hyper-Eddington accretion rates, involves feedback from the BHs altering the conditions on the scale of $R_H$. 
Including the radiative feedback of the BHs on the AGN disk will heat up the CSDs and possibly drive outflows if the radiation pressure is strong enough \citep{Jiang.2014}.
This can limit the BBH accretion rate and weaken the CSD torques \citep{Li.2022}. 
In addition, there is likely also mechanical feedback in the form of a jet powered by accretion onto the BHs on the horizon scale \citep{Paschalidis.2021}. 
If these jets are powerful enough to reach the scales of $R_H$, they may also strongly perturb the CSDs and provide a self-regulation mechanism to BH accretion \citep{Tagawa.2022}. 
Future work on embedded BHs should thus be fully 3D radiation (magneto-)hydrodynamic simulations, and ideally connect the horizon scale to the AGN disk scale.

\begin{acknowledgments}
We thank the referee for a very thorough and helpful report that greatly improved the manuscript. 
AMD gratefully thanks Mathew Bate,  Alex Dittmann, K.E. Saavik Ford, Zoltan Haiman, Dong Lai,  Yaping Li, Jiaru Li,  Barry McKernan,  Diego Mu\~{n}oz, and Judit Szul{\'a}gyi  for helpful conversations and comments. 
We gratefully acknowledge the support by LANL/LDRD under project number 20220087DR.
This research used resources provided by the Los Alamos National Laboratory Institutional Computing Program, which is supported by the U.S. Department of Energy National Nuclear Security Administration under Contract No. 89233218CNA000001.
The LA-UR number is LA-UR-22-22281.
\end{acknowledgments}

\begin{appendix}

\section{Accretion} \label{sec:app_acc}

In this section we outline the mass removal scheme that we apply around each BH in Athena++. 
As emphasized in \citet{Dempsey.2020}, whenever the sink radius is much larger than the physical accretion surface, care must be taken to insure that excess angular momentum not be removed as mass is removed.
This is especially true in Keplerian disks, as the specific angular momentum increases with distance as $\propto \sqrt{r}$.
Because we cannot simulate the true accretion surface of each BH, we opt to conserve the gas angular momentum about the BH as we remove its mass. 

Following the torque-free sink method laid out in \citet{Dempsey.2020} (see also \citealt{Dittmann.2021}), during each time step, if a computational cell is within the sink radius of a BH, and if the gas in that cell is bound to the BH (e.g., the binding energy is negative), we reduce its mass from $m$ to $m'$ where, 
\be \label{eq:app_acc_m}
m' = \frac{m}{1+\gamma n_b \Delta t} ,
\ee
and where $\gamma$ is a parameter. 
Along with the mass, we also change the cell's velocity from $\vec{v}$ to $\vec{v}'$ where,
\be \label{eq:app_acc_v}
\vec{v}' = \vec{v} + \left(\frac{ (\gamma - \beta) n_b \Delta t}{1+\beta n_b \Delta t}\right) \left( \Delta v_\vartheta \uvec{\vartheta} + \Delta v_\varphi \uvec{\varphi} \right) ,
\ee
and where $\beta$ is another parameter, $(\vartheta, \varphi)$ are the angles of a 3D spherical coordinate system centered on the BH, and the velocities are relative to the BH, e.g., $\Delta v_\theta = (\vec{v}- \dot{\vec{r}}_{\rm bh})\cdot \uvec{\vartheta}$.

The parameters $\gamma$ and $\beta$ dictate how much mass is removed each time step and whether the gas conserves ($\beta=0$) or loses ($\beta >0$) its angular momentum about the BH in the process. 
In all of our simulations we set $\beta=0$ appropriate for torque-free mass removal. 
In practice, this means that gas within the sink radius receives a kick in the local $\varphi$ and $\vartheta$ directions. 

We have found through testing that sometimes the fluid may receive very large kicks in the vertical direction during accretion.
This can severely limit the time step, and so we reduce the local value of $\gamma$ whenever the mass or velocity try to change by more than $50\%$ in one time step, i.e. we enforce,
\be
\gamma = \min\left\{ \gamma,  \frac{0.5 |v_i|}{|v_i - \dot{r}_{\rm i,bh}|} , 1\right\} .
\ee
Additionally, we only accrete mass from a cell if the gas in that cell is bound to the BH.
In all of our simulations we set $\gamma$ to an order unity number between $1$ and $4$ depending on the binary separation.

 \end{appendix}

\end{document}